\documentclass[%
 reprint,
superscriptaddress,
 amsmath,amssymb,
 aps,
 pra,
]{revtex4-2}

\usepackage{graphicx}
\usepackage{dcolumn}
\usepackage{bm}
\usepackage{hyperref}
\usepackage{braket}
\usepackage{xcolor}
\usepackage{subfigure}
\usepackage{mathtools}

\usepackage{physics}
\usepackage{soul}

\usepackage{lipsum} 

\begin{document}


\title{Quantum error cancellation in photonic systems - undoing photon losses}

\author{Adam Taylor}
 \email{adam.taylor18@imperial.ac.uk}
 \affiliation{Blackett Laboratory, Imperial College London, London SW7 2AZ, United Kingdom}
\author{Gabriele Bressanini}
\affiliation{Blackett Laboratory, Imperial College London, London SW7 2AZ, United Kingdom}
\author{Hyukjoon Kwon}
\affiliation{Korea Institute for Advanced Study, Seoul 02455, South Korea}
\author{M. S. Kim}
\affiliation{Blackett Laboratory, Imperial College London, London SW7 2AZ, United Kingdom}
\affiliation{Korea Institute for Advanced Study, Seoul 02455, South Korea}


\begin{abstract}
    Real photonic devices are subject to photon losses that can decohere quantum information encoded in the system. In the absence of full fault tolerance, quantum error mitigation techniques have been introduced to help manage errors in noisy quantum devices. In this work, we introduce an error mitigation protocol inspired by probabilistic error cancellation (a popular error mitigation technique in discrete variable systems) for continuous variable systems. We show that our quantum error cancellation protocol can undo photon losses in expectation value estimation tasks. To do this, we analytically derive the (non-physical) inverse photon loss channel and decompose it into a sum over physically realisable channels with potentially negative coefficients. The bias of our ideal expectation value estimator can be made arbitrarily small at the cost of increasing the sampling overhead. The protocol requires a noiseless amplification followed by a series of photon-subtractions. While these operations can be implemented probabilistically, for certain classes of initial state one can avoid the burden of carrying out the amplification and photon-subtractions by leveraging Monte-Carlo methods to give an unbiased estimate of the ideal expectation value. We validate our proposed mitigation protocol by simulating the scheme on squeezed vacuum states, cat states and entangled coherent states. 
\end{abstract}

\maketitle

\section{\label{sec:introduction}Introduction}
Determining whether quantum advantage $-$ defined as the ability for a quantum computer to solve a useful problem significantly faster than any classical method $-$ can be achieved in the near-term constitutes a major open problem. The fragile nature of quantum systems makes them susceptible to errors that could corrupt the computation.
Long-term, we expect that quantum error correction will allow useful quantum algorithms with a proven speedup over their classical counterparts to be implemented reliably \cite{montanaro2016quantum, shor1999polynomial, PhysRevLett.79.325, lloyd1996universal, nielsen2010quantum}.
However, the large hardware overhead and small error threshold requirements \cite{gottesman2010introduction, lidar2013quantum} necessary for quantum error correction to be feasible mean that fault-tolerant universal computation has yet to be achieved.

Currently available quantum devices are characterised by noisy operations and limited number of qubits. \cite{preskill2018quantum, cheng2023noisy, RevModPhys.94.015004, chen2023complexity}. 
Consequently, these noisy devices are suitable to execute limited classes of quantum algorithms, including variational quantum algorithms \cite{cerezo2021variational, tilly2022variational, PhysRevLett.125.010501} and quantum simulations \cite{mccaskey2019quantum, daley2022practical, kim2023evidence}.
In the absence of full fault-tolerance, reducing the impact of noise is an essential ingredient in achieving quantum advantage in the near-term. Quantum error mitigation (QEM) is one potential route to tackling this challenge \cite{cai_quantum_2022, RevModPhys.94.015004, endo2021hybrid}. 

The key insight underlying QEM protocols is that a noise-free expectation value can be extracted from a noisy device by sampling from multiple circuits with suitable post-processing of the noisy measurement data \cite{takagi2022fundamental, quek_exponentially_2023}. Removing unavoidable noise from quantum expectation values is clearly a task with a domain of interest that also extends beyond quantum computing. Therefore, while all these QEM strategies were initially tailored towards the quantum computing community, the techniques may be useful more generally.

QEM techniques have been shown to improve the computational capabilities of noisy devices \cite{kandala2019error, russo2023testing, song2019quantum, berg_probabilistic_2022, kim2023evidence}, with notable examples including zero-noise extrapolation \cite{temme_error_2017, PhysRevX.7.021050} and virtual distillation \cite{PhysRevX.11.041036, PhysRevX.11.031057}. Another promising QEM protocol is probabilistic error cancellation (PEC) \cite{temme_error_2017, PhysRevX.8.031027, piveteau2022quasiprobability, song2019quantum, berg_probabilistic_2022, PRXQuantum.3.010345}. Within this framework, one decomposes the ideal system dynamics into a sum of quantum channels (corresponding to noisy evolutions) with potentially negative coefficients. One then repeatedly probabilistically selects one of these noisy channels, runs the experiment and records the measurement outcomes. These are then appropriately linearly combined to obtain an unbiased estimator of the ideal expectation value. The wide range of applicability and lack of qubit overhead in PEC, combined with the fact it provides an unbiased estimator, make it one of the most promising QEM protocols for potentially achieving quantum advantage in the near-term.

The effectiveness of PEC under a sparse Pauli-Lindblad noise model was demonstrated in Ref.~\cite{berg_probabilistic_2022}, where a sparse learning protocol was proposed that allowed the necessary noise parameters to be accurately estimated. While a significant improvement over unmitigated results was demonstrated, the sampling overhead for this noise model scaled exponentially in qubit number, noise strength and circuit depth.
Building on this work, PEC has recently been extended to circuits that include measurement-based operations such as mid-circuit projective measurements and potential Pauli feedforward operations dependent on those measurement outcomes \cite{gupta2023probabilistic}.

So far, most QEM literature has focused on discrete-variable (DV) quantum systems. 
The continuous-variable (CV) QEM literature \cite{su_error_2021, yang2023post} has mostly been targeted towards Gaussian boson sampling, a non-universal form of quantum computation in which one is interested in sampling from the output photon number distribution obtained by sending squeezed light into a multi-mode passive linear optical interferometer, a task that is believed to be classically intractable under standard complexity theoretic conjectures \cite{v009a004, PhysRevLett.113.100502, hamilton2017gaussian}.
The authors of \cite{su_error_2021} proposed a PEC-inspired protocol to mitigate against the effects of losses in a Gaussian boson sampler.
In particular, their ``loss cancellation’’ scheme does not require changes to the circuit and involves only classical post-processing of the noisy sampling probabilities to infer their ideal counterparts. However, the protocol is limited to photon number detection and its convergence to the noiseless photon sampling distribution is, in general, difficult to assess and only guaranteed in the small-to-intermediate loss regime.
Similarly, the authors of Ref.~\cite{mills2024mitigating} recently introduced a family of strategies for mitigating against losses in DV linear optical quantum computers based around ``recycled probabilities’’.
While their approach outperforms post-selection in the large loss regime, post selection appears to achieve better results for small-to-intermediate losses. It is also limited to DV input states and again to photon number measurements.
\\
\\
In this work, we develop a PEC-inspired scheme to mitigate against photon losses in (either DV or CV) photonic systems in expectation value estimation tasks. To achieve this, we first derive the inverse of the photon loss map.
While such a map is not completely positive (CP)~\cite{nielsen2010quantum}, we show that it can be expressed as a linear combination of physically realisable (i.e., CP) noisy quantum operations.
We then suggest how these operations can be (probabilistically) implemented experimentally. These operations include both an amplification and, counter intuitively, a series of photon subtractions. Sampling from these circuits and applying simple post-processing weights to the measurement outcomes results in an improved estimate of the noise-free ideal expectation value over running the unmitigated circuit. 
The bias of such an estimate can be made arbitrarily small, at the cost of an increasing the sampling overhead.
We also present a Monte-Carlo-based approach that involves sampling from different initial states and linearly combining the measurement outcomes with suitable weights to obtain an unbiased estimator. This approach scales better to many modes than the first approach and can often be implemented deterministically, but only works for specific classes of input state.
We highlight both these methods with clear examples.
Our main contribution is an analytic form of the inverse photon loss channel, as well as possible methods to implement this channel on average, and thus be used as a quantum error mitigation technique which we refer to as quantum error cancellation.

This paper is structured as follows. In Sec.~\ref{sec:preliminaries}, we introduce the basics of quantum error mitigation and outline the probabilistic error cancellation framework. We also describe the noise model for photon losses used throughout this paper and argue that, while not universally applicable, it is relevant and accurate for a variety of important experimental setups..
In Sec.~\ref{sec:probabilistic_error_cancellation}, we present our main results by introducing a quantum error cancellation scheme based around a quasi-probability decomposition of the inverse photon loss map. We discuss its implementation, the classical post-processing cost and sampling overhead, and provide theoretical guarantees of accuracy. 
In Sec.~\ref{sec:examples} we numerically investigate the performance of our protocol via paradigmatic examples involving single-mode squeezed vacuum states and cat states. This section considers just a single mode in order to highlight the key ideas behind the proposed mitigation strategy. 
In Sec.~\ref{sec:multiple_modes}, we extend the scheme to the multi-mode setting where we again discuss implementation, post-processing costs, sampling overhead and provide a thorough error analysis. In Sec.~\ref{sec:two-mode_examples_multi}, we provide numerical simulations of our scheme being applied to multi-mode entangled states, namely two-mode squeezed vacua and entangled coherent states. Finally, we offer concluding remarks in Sec.~\ref{sec:discussion}.


\section{\label{sec:preliminaries}Preliminaries}

\subsection{\label{subsec:quantum_error_mitigation}Quantum Error Mitigation}
In this paper, we are concerned with weak quantum error mitigation \cite{quek_exponentially_2023}. Given an initial state $\rho_0$, and a generic ideal unitary evolution $\mathcal{U}_{\textrm{ideal}}$, the aim of weak quantum error mitigation is to estimate the expectation value $\expval{O}_{\textrm{ideal}} = \Tr[O \, \mathcal{U}_{\textrm{ideal}} [\rho_0]]$ for a generic observable $O$. 
Any attempt to perfectly implement the unitary dynamics ultimately fails due to unavoidable interactions between the system and the environment.
As a result, one is only able to implement a noisy evolution, which we denote with $\mathcal{U}_{\textrm{noisy}}$.
Consequently, the noisy dynamics leads to a measured expectation value that reads $\expval{O}_{\textrm{noisy}} = \Tr[O \, \mathcal{U}_{\textrm{noisy}} [\rho_0]]$.
QEM protocols sample from collections of noisy circuits and combine their measurement outcomes with classical post-processing to give a better estimate of $\expval{O}_{\textrm{ideal}}$, with respect to the unmitigated scenario. 

The quality of an estimator $\bar{O}$ of $\expval{O}_{\textrm{ideal}}$ can be assessed by the mean square error (MSE), which is given by
\begin{align}
    \textrm{MSE} [\bar{O}] &= \mathbb{E} [\left(\bar{O} - \expval{O}_{\textrm{ideal}}\right)^2] \\
    &= \left(\textrm{Bias} [\bar{O}]\right)^2 + \textrm{Var} [\bar{O}] \, , 
\end{align}
where $\textrm{Bias} [\bar{O}] = |\mathbb{E} [\bar{O}] - \expval{O}_{\textrm{ideal}}|$ and $\textrm{Var} [\bar{O}] = \mathbb{E} [\bar{O}^2] - \mathbb{E} [\bar{O}]^2$. 
For example, consider taking $N$ measurements of $O$ on the noisy state $\mathcal{U}_{\textrm{noisy}} [\rho_0]$.
The $N$ measurement outcomes, $(O_1^{(\textrm{noisy})}, O_2^{(\textrm{noisy})}, \dots, O_N^{(\textrm{noisy})})$, may be averaged over to give the noisy estimator $\bar{O}_{\textrm{noisy}} = \frac{1}{N} \sum_{i = 1}^N O_i^{(\textrm{noisy})}$.
The variance of that estimator (often called shot noise) reads $\textrm{Var} [\bar{O}_{\textrm{noisy}}] = \frac{1}{N} \textrm{Var} [O_{\textrm{noisy}}]$, where $\textrm{Var} [O_{\textrm{noisy}}] = \expval{O^2}_{\textrm{noisy}} - \expval{O}^2_{\textrm{noisy}}$ is the single shot variance of $O$ with respect to the noisy output state $\mathcal{U}_{\textrm{noisy}} [\rho_0]$. Therefore, in the large $N$ limit, the dominant source of error is the bias, $|\expval{O}_{\textrm{noisy}} - \expval{O}_{\textrm{ideal}}|$.

QEM protocols aim to provide an estimator of $\expval{O}_{\textrm{ideal}}$, denoted with $\bar{O}_{\textrm{mit}}$, with a smaller bias than $\bar{O}_{\textrm{noisy}}$. However, this comes at the cost of an increases variance and hence to obtain a mitigated estimator with the same shot noise as the unmitigated estimator will require more data samples to be taken.
This \emph{sampling overhead} approximately amounts to $\textrm{Var} [\bar{O}_{\textrm{mit}}] / \textrm{Var} [\bar{O}_{\textrm{noisy}}]$ \cite{cai_quantum_2022}.
While the exact details of the bias/sampling overhead trade-off relation depend on the particular mitigation scheme, the sampling overhead typically scales exponentially with noise strength, circuit depth and number of qubits \cite{takagi2022fundamental, quek_exponentially_2023, PhysRevLett.131.210601, PhysRevLett.131.210602}.


\subsubsection{\label{subsubsec:probabilistic_error_cancellation_preliminaries}Probabilistic Error Cancellation}
In Ref.~\cite{temme_error_2017}, Temme \textit{et al} demonstrated that, given a quasi-probability decomposition of the ideal dynamics $\mathcal{U}_{\textrm{ideal}}$ in terms of some arbitrary physically realisable (CP) noisy quantum operations $\{ \mathcal{F}_j \}$ \footnote{The set of noisy operators $\{ \mathcal{F}_j \}$ should form an noisy operator basis into which $\mathcal{U}_{\textrm{ideal}}$ can be decomposed.},
\begin{equation}
\label{eq:quasiprobability_representation_of_ideal_unitary_dynamics/evolution}
    \mathcal{U}_{\textrm{ideal}} = \sum_j \omega_j \, \mathcal{F}_j \, , 
\end{equation}
it is possible to produce an unbiased estimate of $\expval{O}_{\textrm{ideal}}$. Here, $\{\omega_j\}$ is a set of real coefficients that, in order for $\mathcal{U}_{\textrm{ideal}}$ to be trace-preserving, satisfy $\sum_j \omega_j = 1$.
We refer to $\omega_j$ as a quasi-probability distribution and to Eq.~\eqref{eq:quasiprobability_representation_of_ideal_unitary_dynamics/evolution} as a quasi-probability representation of $\mathcal{U}_{\textrm{ideal}}$. 

In practice, given a noise CPTP map, one computes the mathematical inverse of such a map (assuming such an inverse exists).
This is typically a non-CP map and hence cannot be physically realised. 
However, the inverse map always admits a quasi-probability representation in terms of physically realisable quantum operations. This can then be combined with the full circuit dynamics to write the ideal circuit dynamics in the quasi-probability representation over noisy CP channels. We refer to Ref.~\cite{berg_probabilistic_2022} for a detailed demonstration of how one can obtain a quasi-probability decomposition of the ideal dynamics, starting from the noise model. Note that by doing this, one is implicitly assuming that the noise channel and ideal dynamics can be separated (see, e.g.; \cite{PhysRevApplied.15.034026} for a detailed discussion on this point).

We then use the quasi-probability distribution $\{ \omega_j \}$ to define a new set of coefficients $q_j := |\omega_j| / S$, where $S = \sum_j |\omega_j|$. These coefficients satisfy $0 \le q_j \le 1$ and $\sum_j q_j = 1$, and hence form a valid probability distribution. The quasi-probability representation of $\mathcal{U}_{\textrm{ideal}}$ can then be rewritten as
\begin{equation}
    \mathcal{U}_{\textrm{ideal}} = S \sum_j \frac{|\omega_j|}{S} \textrm{sgn} (\omega_j) \mathcal{F}_j.
\end{equation} 
Via linearity, it is clear that
\begin{equation}
    \expval{O}_{\textrm{ideal}} = S \sum_j q_j \, \textrm{sgn}(\omega_j) \expval{O}_{\mathcal{F}_{j}} \, ,
\end{equation}
where $\expval{O}_{\mathcal{F}_j} = \Tr\left[ O \, \mathcal{F}_j [\rho_0] \right]$.
Using a Monte Carlo approach, the ideal expectation value can be estimated by sampling from the channels $\{\mathcal{F}_j \}$ according to the probability distribution $\{q_j \}$, and performing a single shot measurement of $O$ - this probabilistic choice of $\mathcal{F}_j$ is where probabilistic error cancellation gets its name. We thus obtain a list of measurement outcomes $(O_1^{(\mathcal{F}_{j_1})}, O_2^{(\mathcal{F}_{j_2})}, \dots, O_N^{(\mathcal{F}_{j_N})})$, where the superscript $\left(\mathcal{F}_{j_\ell}\right)$ tells us which channel was implemented in the $\ell^{\textrm{th}}$ run of the experiment. 
These $N$ samples may then be linearly combined to obtain an \emph{unbiased} mitigated estimator, namely
\begin{equation}
    \bar{O}_{\textrm{mit}} = \frac{S}{N} \sum_{\ell = 1}^N \textrm{sgn}(\omega_{j_\ell}) O_\ell^{\mathcal{F}_{j_\ell}}.
\end{equation}
Under the assumption that $\textrm{Var} [O_{\mathcal{F}_j}] / \textrm{Var} [O_{\textrm{noisy}}] \approx \mathcal{O}(1)$ for all $j$, the sampling overhead $\textrm{Var} [\bar{O}_{\textrm{mit}}] / \textrm{Var} [\bar{O}_{\textrm{noisy}}]$ can easily be shown to scale with $S^2$.


\subsection{\label{subsec:noise_model}Noise Model}

\begin{figure}
    \centering
    \includegraphics[width = 0.45\textwidth]{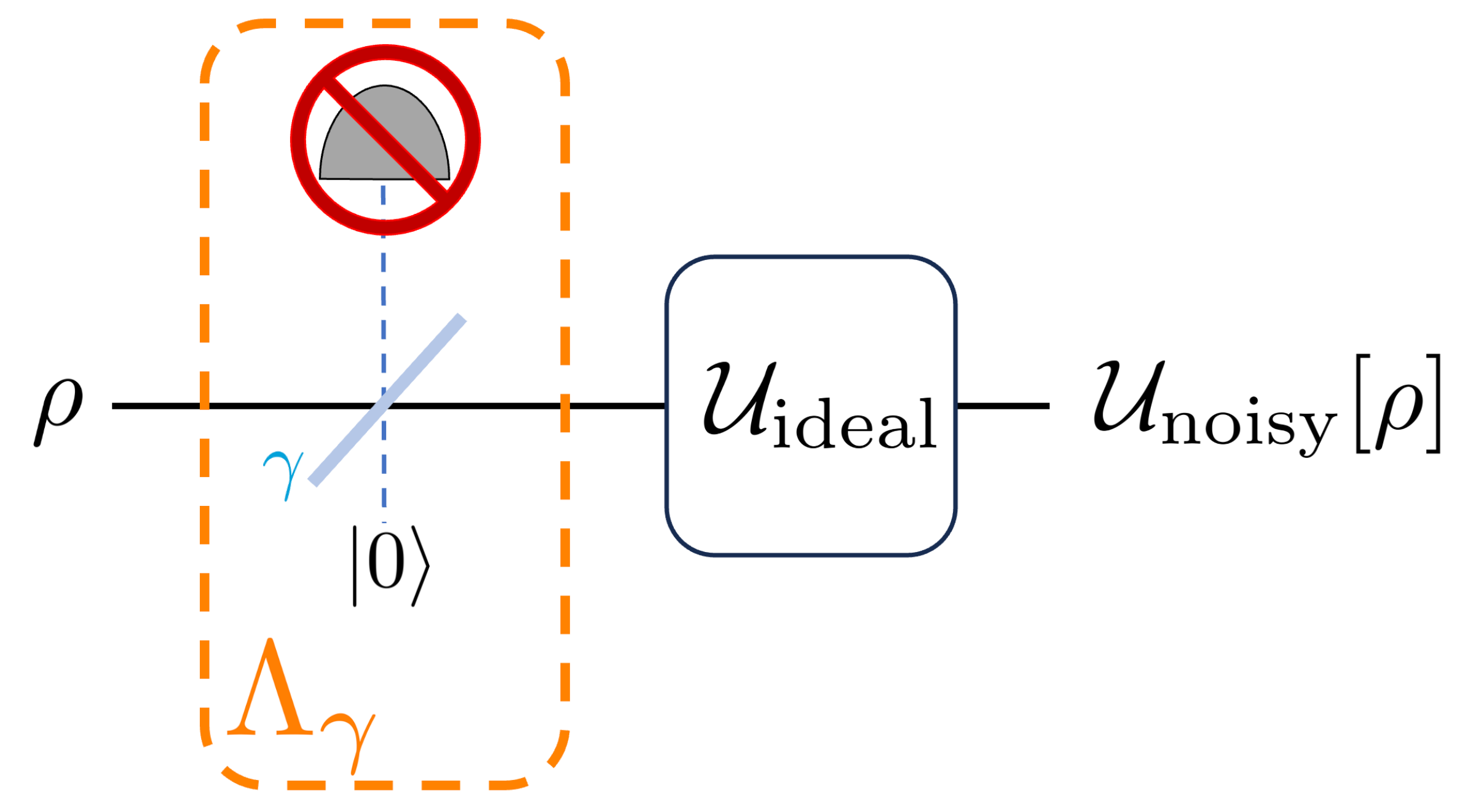}
    \caption{Graphical representation of the decomposition of the noisy dynamics into a pure loss map followed by an ideal unitary evolution for a single mode, i.e., $\mathcal{U}_{\textrm{noisy}} = \mathcal{U}_{\textrm{ideal}} \circ \Lambda_\gamma$.
    The loss map is modeled via a beam splitter interaction between the system and an environmental mode initialized in the vacuum state, followed by a trace over the environmental degrees of freedom.
    The beam splitter has a transmissivity of $1 - \gamma$, where $\gamma \in [0, 1)$ is the loss parameter.}
    \label{fig:noise_model_diagram}
\end{figure}

In photonic platforms, the two dominant sources of noise are photon losses and dephasing errors, while thermal noise is negligible for systems at visible or infrared  frequencies near room temperature. In this work, we focus on the photon loss channel because it is the most significant issue impacting photonic platforms. In Appendix~\ref{dephasing_error_section}, we discuss both a strategy for mitigating against dephasing errors and also show that the introduction of a dephasing error channel has no impact on our photon loss mitigation procedure.

Let $\Lambda_\gamma$ denote the single-mode pure Markovian photon loss channel with noise parameter $\gamma \in [0, 1]$. We model these losses via a beam-splitter acting on a single-mode input state $\rho$ and an ancillary vacuum state, before tracing over the environmental degrees of freedom, namely \cite{PhysRevA.52.2401}
\begin{equation}\label{eq:photon_loss_beam_splitter_stinespring_dilation_representation}
    \Lambda_{\gamma} [\rho] = \Tr_{\textrm{env}}[ U_{\textrm{BS}, \gamma} \, \left(\rho \otimes \ketbra{0} \right) \, U^\dag_{\textrm{BS}, \gamma}].
\end{equation}
Here, $U_{\textrm{BS},\gamma} = e^{\textrm{arccos}(\sqrt{1 - \gamma})(\hat{a}_{\textrm{env}}^\dag \hat{a} - \hat{a}^\dag \hat{a}_{\textrm{env}})}$
denotes the beam splitter unitary, and $1-\gamma$ its trasmissivity, while $\hat{a}$ and $\hat{a}_{\textrm{env}}$ are the bosonic operators of the system and environment, respectively.
For future convenience, we exploit the state's Glauber-Sudarshan $P$-representation \cite{glauber1963quantum}
\begin{equation}
    \label{eq:P_function_definition}
    \rho = \int_\mathbb{C} \textrm{d}^2\alpha \, P_\rho(\alpha) \ketbra{\alpha} \, ,
\end{equation}
where $\ket{\alpha}$ is a single-mode coherent state and $P_\rho$ is the $P$ function of $\rho$.
Hence, we may express Eq.~\eqref{eq:photon_loss_beam_splitter_stinespring_dilation_representation} as
\begin{equation}\label{eq:photon_loss_P_representation}
    \Lambda_\gamma[\rho] = \int_{\mathbb{C}} \textrm{d}^2 \alpha P_\rho(\alpha) \ketbra{\sqrt{1 - \gamma} \, \alpha} \, .
\end{equation}
One can also use Eq.~\eqref{eq:photon_loss_beam_splitter_stinespring_dilation_representation} to obtain a Kraus representation of the loss map, namely
\begin{equation}\label{eq:photon_loss_kraus_representation}
    \Lambda_\gamma[\rho] = \sum_{j=0}^\infty K_j(\gamma) \, \rho \, K_j^\dag(\gamma) \, , 
\end{equation}
where the Kraus operators read
\begin{align}
\begin{split}
\label{eq:Kraus_operators}
    K_j(\gamma) &= \sqrt{\frac{\gamma^j}{j!}}(1-\gamma)^{\frac{\hat{n}}{2}}\hat{a}^j \\
    &= \sqrt{\frac{1}{j!} \left(\frac{\gamma}{1 - \gamma}\right)^j} \hat{a}^j (1 - \gamma)^{\frac{\hat{n}}{2}} \, .
\end{split}
\end{align}
We prove the above identity above in Appendix~\ref{appendix_sec:useful_commutation_relation}. As the channel is trace-preserving, the Kraus operators satisfy $\sum_j K_j^\dag K_j = \mathbb{I}$, where $\mathbb{I}$ denotes the identity operator on the Hilbert space. 
Equivalently, it can be seen that Eq.~\eqref{eq:photon_loss_beam_splitter_stinespring_dilation_representation} is a solution to the Lindblad master equation 
\begin{equation}
\label{eq:photon_loss_lindblad_master_equation_representation}
    \frac{d}{dt} \rho (t) = \kappa \mathcal{D}[\hat{a}](\rho(t)) \, , 
\end{equation}
where $\mathcal{D}[\hat{a}](\rho) = \hat{a} \rho \hat{a}^\dag - \frac{1}{2} \{\hat{a}^\dag \hat{a},\, \rho\}$ is the dissipation superoperator and $\kappa > 0$ is the dissipation rate, if we make the identification $\gamma = 1 - e^{-\kappa t}$.

Crucially, we further assume that the noisy dynamics $\mathcal{U}_{\textrm{noisy}}$ can be decomposed into a pure loss map taking place \emph{before} the system undergoes an ideal evolution, i.e., we write $\mathcal{U}_{\textrm{noisy}} = \mathcal{U}_{\textrm{ideal}} \circ \Lambda_\gamma$, as depicted in Fig.~\ref{fig:noise_model_diagram}. A more precise modelling of the noisy dynamics would necessitate solving the full Lindblad master equation in Eq.~\eqref{eq:photon_loss_lindblad_master_equation_representation} with an additional term on the right-hand-side corresponding to the ideal dynamics.
Nonetheless, while the assumption above is not universally applicable, we point out that it is generally justified in a number of relevant cases, particularly in the multi-mode setting, which we discuss below.

We now extend the above noise model to multiple modes, under the assumption that each mode is subject to independent single-mode Markovian photon losses parameterised by some learnable loss parameters. For $M$-modes, let $\bm{\gamma} = (\gamma_1, \gamma_2, \dots \gamma_M)$ denote these loss parameters in each mode. 
The full $M$-mode loss channel, $\Lambda_{\bm{\gamma}}^{(M)}$, can then be written as a tensor product over the single mode loss channels, i.e.,
\begin{equation}\label{eq_mm_loss_channel}
    \Lambda_{\bm{\gamma}}^{(M)} = \bigotimes_{i = 1}^M \Lambda_{\gamma_i, i}
\end{equation}
where $\Lambda_{\gamma_i, i}$ only acts non-trivially on the $i^{\textrm{th}}$ mode. In the $P$-representation, an $M$-mode state can be written as 
\begin{equation}
    \rho = \int_{\mathbb{C}^M} \dd[2M]\bm{\alpha} P_{\rho} (\bm{\alpha}) \ketbra{\alpha_1, \dots, \alpha_M},
\end{equation}
where $\bm{\alpha} = (\alpha_1, \alpha_2, \dots, \alpha_M)$ and $\ket{\alpha_1, \dots, \alpha_M} = \bigotimes_{i = 1}^M \ket{\alpha_i}$. Photon losses then take the familiar form
\begin{align}\label{mm_loss_channel_P_representation}
\begin{split}
    \Lambda_{\bm{\gamma}}^{(M)} [\rho] = \int_{\mathbb{C}^M} \dd{\bm{\alpha}} P_\rho (\bm{\alpha}) \ketbra{\sqrt{1 - \gamma_1} \alpha_1} \otimes \\
\dots \otimes \ketbra{\sqrt{1 - \gamma_M} \alpha_M} .
\end{split}
\end{align}
It can be seen that uniform losses commute with linear optical elements \cite{oszmaniec2018classical}. In this case, losses occurring at any point in the dynamics (including uniform detector inefficiencies) can be ``shifted'' to the front of the circuit.
Therefore, a noisy passive linear optical network in which each path is subject to approximately the same level of loss is well modelled by $\mathcal{U}_{\textrm{noisy}} = \mathcal{U}_{\textrm{ideal}} \circ \Lambda_{\bm{\gamma}}^{(M)}$. 

In Ref.~\cite{clements2016optimal}, Clements \textit{et al} provided an optimal decomposition of any $M-$mode unitary $U \in SU(2M)$ using only passive linear optical components.
Their decomposition is \emph{balanced} because each path within the interferometer passes through the same number of beam splitters. This in turn implies that photon loses are well modelled as being uniform thus matching our noise model.

Non-uniform (i.e., path-dependent) losses can not be perfectly separated from the ideal passive linear optical dynamics. However, the authors of Ref.~\cite{brod2020classical} showed that \emph{some} losses can still  be shifted to the front of the circuit, leaving behind a \emph{new} noisy passive linear optical network (described by $\mathcal{U}'_{\textrm{noisy}}$) with less noise, namely $\mathcal{U}_{\textrm{noisy}} = \mathcal{U}'_{\textrm{noisy}} \circ \Lambda_{\bm{\gamma'}}^{(M)}$.
In principle, one could then use our protocol to partially mitigate the effect of losses, i.e., those coming from $\Lambda_{\bm{\gamma'}}^{(M)}$. However, we point out that determining the individual loss parameters $(\gamma_1, \gamma_2, \dots, \gamma_M)$ and the exact form of the new $\mathcal{U}'_{\textrm{noisy}}$ may be challenging.

In realistic settings, one would expect the loss parameter to fluctuate slightly between shots. This can be modelled by selecting $\gamma$ from a Gaussian distribution centred on $\bar{\bm{\gamma}} := \mathbb{E}[\gamma]$ with standard deviation $\bm{\sigma_\gamma}$. Over longer time periods, the average loss parameter $\bar{\bm{\gamma}}$ could vary and hence one should periodically re-measure the loss parameter. In practise however, the high shot rate associated with photonic platforms means we do not expect $\bar{\bm{\gamma}}$ to appreciably vary over the timescale of a single experimental run.
\\
\\
While our discussion is focused on photonic platforms, our results are applicable to generic bosonic systems described by the canonical commutation relations $[\hat{a}_i, \hat{a}_j^\dag] = \delta_{ij} \mathbb{I}$.
Consequently, our mitigation protocol can be applied in a diverse range of bosonic architectures.

\subsubsection{Learning the Loss Parameter}

The average loss parameter can be estimated by inputting a multi-mode coherent state $\ket{\bm{\alpha}} = \ket{\alpha_1, \alpha_2, \dots, \alpha_M}$ and measuring the output intensity, under the assumption of uniform losses, noisy dynamics that can be decomposed as in Fig.~\ref{fig:noise_model_diagram} and passive ideal dynamics. The input intensity is proportional to the expected number of input photons,  $\langle \hat{N} \rangle_{\textrm{in}} = \sum_{i = 1}^M |\alpha_i|^2$, where $\hat{N} = \sum_{i = 1}^M \hat{n}_i$. The multi-mode photon loss channel transforms these coherent states into $\ket{\bm{\alpha}} \xrightarrow{\Lambda_{\bm{\gamma}}} \ket{\sqrt{1 - \gamma} \alpha_1, \, \sqrt{1 - \gamma} \alpha_2, \, \dots, \sqrt{1 - \gamma} \alpha_M}$. Via energy conservation, the expected output number of photons will therefore be given by $\langle\hat{N}\rangle_{\textrm{out}} = \Tr[\hat{N} \, \mathcal{U}_{\textrm{ideal}} \circ \Lambda_{\bm{\gamma}} [\ketbra{\bm{\alpha}}]] = \sum_{i = 1}^M |\sqrt{1 - \gamma} \alpha_i|^2 = (1 - \gamma) \langle\hat{N}\rangle_{\textrm{in}}$. Denoting our estimate of $\gamma$ as $\tilde{\gamma}$, it is given 
\begin{align}
\begin{split}
    \tilde{\gamma} &= 1 - \frac{\langle\hat{N}\rangle_{\textrm{out}}}{\langle\hat{N}\rangle_{\textrm{in}}} \\
    &=1 - \frac{I_{\textrm{out}}}{I_{\textrm{in}}}
\end{split}
\end{align}
where $I_{\textrm{in/out}}$ is the average total intensity summed across all modes going in/out of the circuit. 
Due to the fact that coherent states are well approximated by laser fields, which are among the most stable states one can prepare in quantum optical experiments, this reduces the impact of state preparation errors on our estimate of $\gamma$.
State preparation errors can be further alleviated by varying across a variety of different choices $\ket{\bm{\alpha}}$ and taking the average estimated loss parameter. Inefficient detectors in quantum optics are modelled as a photon loss occurring immediately before detection, and can there be incorporated into our estimate of $\gamma$ (and hence mitigated against by our error cancellation protocol).

If $\mathcal{U}_{\textrm{ideal}}$ is a passive linear optical network, then the output state will be the multi-mode coherent state $\ket{\bm{\beta}} = \ket{\beta_1, \beta_2, \dots, \beta_M}$. 
When estimating the output intensity by making $N_{\textrm{shots}}$ measurements, the variance of $\tilde{\gamma}$ will be
\begin{align}
\begin{split}
    \textrm{Var}[\tilde{\gamma}] &= \frac{1}{N_{\textrm{shots}}} \textrm{Var} \left[1 - \frac{I_{\textrm{out}}}{I_{\textrm{in}}}\right] \\
    &= \frac{1}{N_{\textrm{shots}}} \textrm{Var} \left[\frac{\langle \hat{N} \rangle_{\textrm{out}}}{\langle \hat{N} \rangle_{\textrm{in}}} \right] \\
    &=\frac{1}{N_{\textrm{shots}} \langle \hat{N} \rangle_{\textrm{in}}^2} \textrm{Var} \sum_{i = 1}^M \expval{n_i}_{\textrm{out}} \\
    &= \frac{(1 - \gamma)^2}{N_{\textrm{shots}} \sum_i |\alpha_i|^2}.
\end{split}
\end{align}
One can reduce this variance by increasing the number of shots or increasing the total intensity of the initial coherent state $\ket{\bm{\alpha}}$. This also provides guarantees about the level of accuracy via Chebyshev's inequality; for example, to have a $>99\%$ probability of estimating $\gamma$ to within an accuracy $< 0.01$ when $\sum_{i = 1}^M |\alpha_i|^2 = M$ requires only $N_{\textrm{shots}} > 1,000,000 / M$. Given that coherent states can be easily produced, this number of shots can be achieved in less than $1$ second for many optical platforms.

Alternatively, for passive linear optical networks one could incorporate learning the losses with learning the exact ideal unitary being implemented by using one of the methods proposed in Refs.~\cite{rahimi2013direct, dhand2016accurate}. This has the advantage of also giving information about exactly what $\mathcal{U}_{\textrm{ideal}}$ is, but is a more involved process.


\section{\label{sec:probabilistic_error_cancellation}Quantum Error Cancellation}
In this section, our proposed quantum error cancellation scheme is presented for a single mode.
\begin{figure}
    \centering
    \includegraphics[width = 0.45\textwidth]{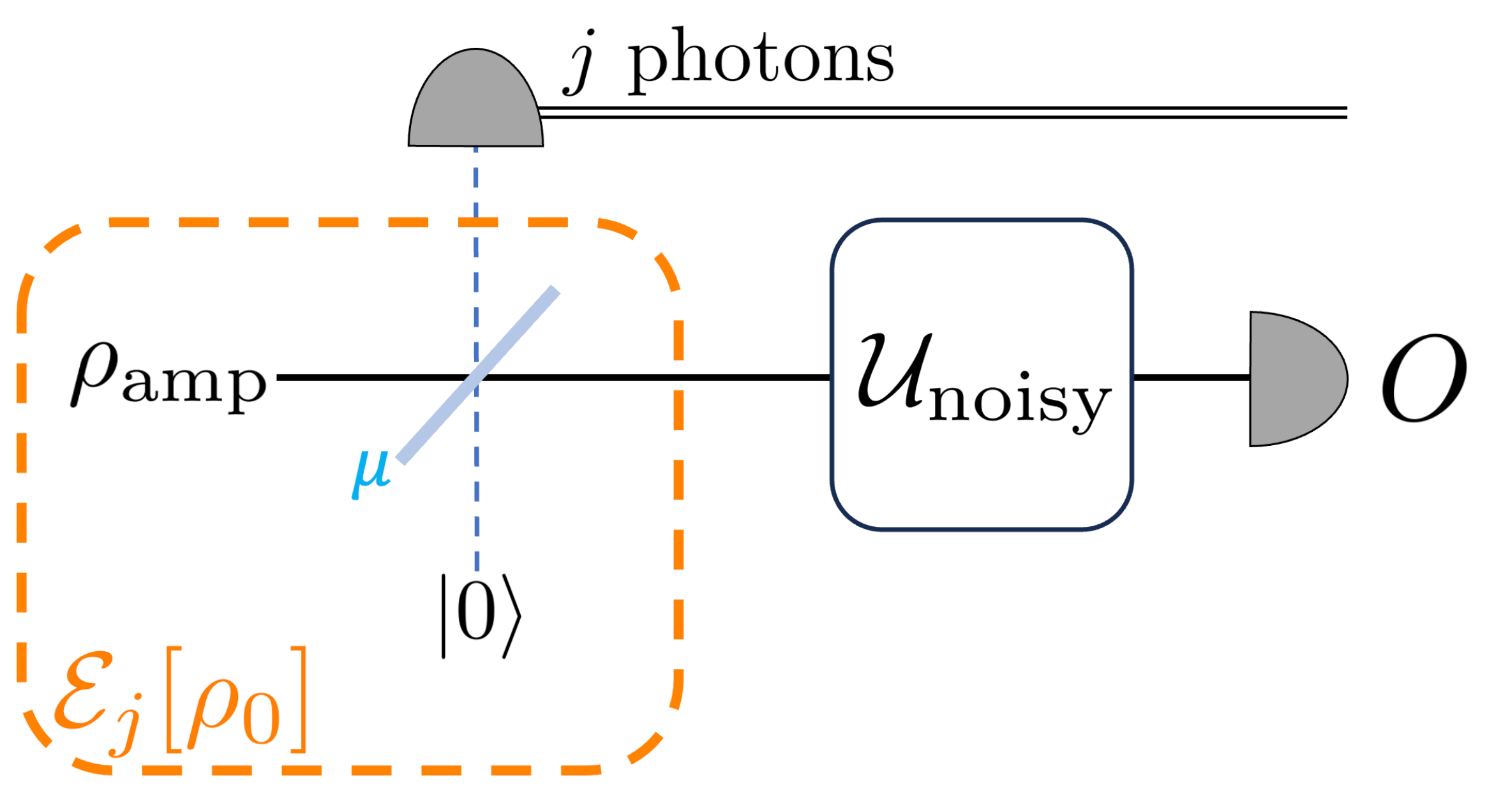}
    \caption{Schematic diagram depicting the full circuit used in the single-mode mitigation scheme as described in Sec.~\ref{subsec:probabilistic_error_cancellation}.
    If one wants to estimate the noise-free expectation value $\expval{O}_{\textrm{ideal}} = \Tr[O \, \mathcal{U}_{\textrm{ideal}} [\rho_0]]$, one first prepares the amplified state $\rho_{\textrm{amp}} \sim g_{\mu}^{\hat{n}} \rho_0 g_{\mu}^{\hat{n}}$. This state is sent through the depicted beam splitter with a vacuum in the ancilla input port. A measurement of $j$ photons from the ancilla output port will herald the state $\mathcal{E}_j [\rho_0]$ - this is as if we physically started with $\rho_0$ and implemented the CP channel $\mathcal{E}_j$. Which channel is implemented in each individual shot is determined by the number of detected photons $j$. Following this, the noisy dynamics $\mathcal{U}_{\textrm{noisy}}$ are run and a measurement of $O$ is made. This circuit is run many times to obtain estimates of $\expval{O}_{j, \textrm{noisy}} = \Tr[O \, (\mathcal{U}_{\textrm{noisy}} \circ \mathcal{E}_j) [\rho_0]]$ for different values of $j$. These results are then linearly combined with the relevant coefficients $\omega_j (\rho_0)$ to give the mitigated estimator $\bar{O}_{\textrm{mit}}$. All interventions take place before the state undergoes its noisy dynamics and so it is not necessary to reconfigure or change the circuit between experimental shots.}
    \label{fig:single_mode_channel_selection/full_circuit_diagram}
\end{figure}

\subsection{\label{subsec:inverse_photon_loss}Inverse Loss Map}
In this section, we construct the inverse $\Lambda_\gamma^{-1}$ to the pure loss map.
In particular, Eq.~\eqref{eq:photon_loss_P_representation} implies that the inverse photon loss map, expressed using the $P$ representation, must read
\begin{equation}\label{eq:inverse_photon_loss_P_representation}
    \Lambda_\gamma^{-1} [\rho] = \int_{\mathbb{C}} \textrm{d}^2 \alpha \, P_\rho(\alpha) \ketbra{g_0 \alpha} \, ,
\end{equation}
where $g_0 = 1 / \sqrt{1 - \gamma} > 1$.
This immediately leads to
$\Lambda_{\gamma} \left[ \Lambda_\gamma^{-1} [\rho] \right] = \Lambda_\gamma^{-1} \left[ \Lambda_\gamma [\rho] \right] = \rho$. 
In other words, $\Lambda_\gamma^{-1}$ effectively implements a noiseless amplification on every constituent coherent state in the $P$-representation \cite{ralph2009nondeterministic} \emph{without} changing their relative weights.

An operator-sum representation of this inverse loss map is given by
\begin{equation}\label{eq:inverse_photon_loss_operator_sum_representation}
    \Lambda_\gamma^{-1} [\rho] = \sum_{j = 0}^\infty \frac{(-\gamma)^j}{j!} \, \hat{a}^j g_0^{\hat{n}} \, \rho \, g_0^{\hat{n}} (\hat{a}^{\dag})^j.
    \end{equation}
The equivalence between  Eq.~\eqref{eq:inverse_photon_loss_P_representation} and Eq.~\eqref{eq:inverse_photon_loss_operator_sum_representation} is shown in Appendix~\ref{appendix_sec:inverse_excitation_loss_proof}. 
It can be easily seen that $\Lambda_\gamma^{-1}$ is Hermitian-preserving and trace-preserving, but not completely positive (as evident by the $(-\gamma)^j$ term), and hence is non-physical.
Nevertheless, Eq.~\eqref{eq:inverse_photon_loss_operator_sum_representation} may be expressed as a linear combination of CP maps $\lbrace\mathcal{E}_j\rbrace$, namely
\begin{equation}\label{eq:inverse_photon_loss_decomposition_physcially_realizable_channels}
    \Lambda_{\gamma}^{-1} [\rho] = \sum_{j = 0}^{\infty} \omega_{j}(\rho) \mathcal{E}_j [\rho] \, ,
\end{equation}
where 
\begin{equation}
    \mathcal{E}_j [\rho] = \frac{1}{\mathcal{N}_{j}(\rho)} \hat{a}^j g_0^{\hat{n}} \, \rho \, g_0^{\hat{n}} (\hat{a}^\dag)^j \, ,
    \label{eq:noisy_state_j}
\end{equation}
with
\begin{equation}
    \label{eq:E_j_normalisaion_definition}
    \mathcal{N}_{j}(\rho) = \Tr[\hat{a}^j g_0^{\hat{n}} \, \rho \, g_0^{\hat{n}} (\hat{a}^\dag)^j] \, , 
\end{equation}
and the (possibly negative) coefficients are given by 
\begin{equation}
    \label{eq:omega_coefficient_definition}
    \omega_{j}(\rho) = \frac{(-\gamma)^j}{j!} \mathcal{N}_j(\rho) \, .
\end{equation}
Note that while these maps are not trace preserving, we will use a slight abuse of notation and refer to them as channels.

These channels can be, at least \emph{probabilistically}, physically realised. 
Subject to the noise model assumptions discussed in the previous section, we conclude that any ideal system evolution may be decomposed into a weighted sum of noisy, physically-realisable channels, i.e.,
\begin{equation}
    \label{eq:cv_ideal_dynamics_quasiprobability}
    \mathcal{U}_{\textrm{ideal}} [\rho] = \sum_j \omega_j (\rho) (\mathcal{U}_{\textrm{noisy}} \circ \mathcal{E}_j) [\rho] \, .
\end{equation}
By making the identification $\mathcal{F}_j = (\mathcal{U}_{\textrm{noisy}} \circ \mathcal{E}_j)$, it can be seen that this is a quasi-probability representation of the ideal dynamics  Eq.~\eqref{eq:quasiprobability_representation_of_ideal_unitary_dynamics/evolution}.

We point out that for certain states  Eq.~\eqref{eq:inverse_photon_loss_operator_sum_representation} can be rewritten to contain only a finite number of terms. For example, for states containing at most $K - 1$ photons, it is clear that only the first $K$ terms of the operator-sum representation of $\Lambda_\gamma^{-1}$ act non-trivially. Similarly, if $\rho = \ketbra{\psi}$ and $g^{\hat{n}} \ket{\psi}$ is an eigenstate of $\hat{a}^K$, then Eq.~\eqref{eq:inverse_photon_loss_operator_sum_representation} can be rewritten such that it contains only $K$ terms.

\subsubsection{\label{subsubsec:fock_intuition_for_loss}Intuition Behind Inverse Photon Loss Map}
We now aim to briefly provide some physical intuition
behind the $-$ somewhat unexpected $-$ presence of photon-subtracting terms $\hat{a}^j$ in the inverse of the loss map Eq.~\eqref{eq:inverse_photon_loss_operator_sum_representation}.
For simplicity, let us consider the photon loss map and its inverse acting on the Fock state $\ketbra{m}$, i.e.,
\begin{equation}
\label{eq:physical_intuition_for_inverse_loss_channel_lambda_lambda_-1}
\begin{split}
    & (\Lambda_\gamma \circ \Lambda_\gamma^{-1}) [\ketbra{m}] =  \\ & 
    \sum_{i = 0}^\infty \sum_{j = 0}^\infty \frac{1}{i!}\frac{1}{j!} \left(\frac{\gamma}{1 - \gamma}\right)^i \left(\frac{-\gamma}{1 - \gamma}\right)^j  \hat{a}^{i + j} 
    \ketbra{m}
    (\hat{a}^\dag)^{i + j} \, .
\end{split}
\end{equation}
We can see that the noiseless amplification term $g^{\hat{n}}$ appearing in $\Lambda_\gamma^{-1}$ cancels out with the $g^{-\hat{n}}$ term present in the loss map expansion, leaving behind a series of photon subtraction terms only.
Using the fact that at most $m$ photons may be subtracted from $\ketbra{m}$, one can show that Eq.~\eqref{eq:physical_intuition_for_inverse_loss_channel_lambda_lambda_-1} can be expressed as
\begin{equation}
    \sum_{n=0}^m \left[ \sum_{j=0}^{m - n} (-1)^j \binom{m-n}{j} \right] \binom{m}{n} \left( \frac{\gamma}{1-\gamma}\right)^{m-n} \ketbra{n}\, .
\end{equation}
For $n < m$, the alternating sign causes the coefficient inside the square brackets to vanish.
We remind the reader that the alternating sign originates from the operator-sum representation of the inverse loss map.
On the other hand, for $n = m$, said coefficient is equal to 1 and  hence $(\Lambda_\gamma \circ \Lambda_\gamma^{-1}) [\ketbra{m}] = \ketbra{m}$ as expected.

Physically, we can understand these cancellations in terms of quantum trajectories. 
If we consider just the photon loss map, then there is clearly only one trajectory in which the input Fock state loses $k$ photons (i.e., via the action of the Kraus operator $K_k$).
On the other hand, we can see from Eq.~\eqref{eq:physical_intuition_for_inverse_loss_channel_lambda_lambda_-1} that the presence of the inverse loss maps allows the existence of \emph{multiple} trajectories in which the system loses $k > 0$ photons, namely every $i$ and $j$ in Eq.~\eqref{eq:physical_intuition_for_inverse_loss_channel_lambda_lambda_-1} such that $i + j = k$.
The alternating sign in Eq.~\eqref{eq:inverse_photon_loss_operator_sum_representation} leads to the cancellation of these trajectories.
Conversely, there is only one trajectory in which the system does not lose any photons (i.e., $i + j = 0$), hence in this case the cancellation can not take place and the original noise-free state is returned.


\subsection{\label{subsec:probabilistic_error_cancellation}Quantum Error Cancellation Protocol}
Our mitigation scheme places all interventions before the state undergoes its noisy dynamics $\mathcal{U}_{\textrm{ideal}} \circ \Lambda_\gamma$.

To realise the channels $\lbrace\mathcal{E}_j\rbrace$, one needs to implement an \emph{amplification} operation, $g^{\hat{n}}$, and a $j$-photon subtraction, $\hat{a}^j$.
The operator $g^{\hat{n}}$ is said to implement a
(noiseless) amplification operation because, up to normalisation, it has the effect of amplifying a coherent state's amplitude by a factor of $g>1$. The standard implementation of photon subtraction requires sending the input state through a beam splitter which further attenuates the state \cite{kim2008recent}. 
However, we can compensate for this spurious effect in the amplification step by considering a gain factor $g_\mu > g_0$ that depends on the beam splitter transmissivity. In the following sections we discuss these aspects in detail.


\subsubsection{\label{subsubsec:amplification}Amplification}
Currently available protocols for noiseless amplification are notoriously challenging to implement ~\cite{ralph2009nondeterministic, zavatta2011high, winnel2020generalized, PhysRevLett.128.160501_teleamplification}.
This is primarily due to their inherent probabilistic nature and to their demanding resources requirements, such as multiple single-photon Fock states, that limits scalability to many modes.

One approach by Zavatta \textit{et al} \cite{zavatta2011high} achieves noiseless amplification in the Hilbert subspace spanned by the Fock states $\lbrace\ket{0}, \ket{1}\rbrace $ by using a nonlinear optical element to probabilistcally implement the operator $\mathbb{I} + c \hat{n}$.
Looping this scheme $N$ times with different constants $c_i$ leads to the implementation of the operator $\prod_{i=1}^N (\mathbb{I} + c_i \hat{n})$. By carefully tuning the parameters $c_i$, one can implement exact noiseless amplification within the subspace spanned by $\{\ket{0}, \ket{1}, \dots \ket{N}\}$. 
Alternatively, if the amplification factor $g_{\mu}$ is close to 1, one could use this same approach to implement the truncated Taylor series, $g_\mu^{\hat{n}} \approx \sum_{i = 0}^N \ln(g_\mu)^i \hat{n}^i / i!$. 
Similarly, the teleamplification scheme by Guanzon \emph{et al} \cite{PhysRevLett.128.160501_teleamplification} is a generalised quantum scissor operation that transforms a infinite dimensional state $\ket{\psi} = \sum_{n = 0}^\infty b_n \ket{n}$ into a finite dimensional state $\ket{\psi} \rightarrow \sum_{n = 0}^N g^n b_n \ket{n}$ in the Hilbert space spanned by $\left\{ \ket{0}, \ket{1}, \dots, \ket{N} \right\}$. While this technique does not require non-linear optical elements and has a larger success probability than Zavatta \emph{et al}'s scheme (and in fact asymptotically saturates the maximal possible success probability when amplifying in the $2$-dimensional subspace, see \cite{guanzon2022saturating}), it requires a resource state of many single photon Fock states.

If the amplification operation takes place within the circuit, it could then be advantageous to first do the photon subtraction as the latter operation could force a larger fraction of the state $\rho_0$ into the subspace spanned $\{\ket{0}, \dots, \ket{N}\}$ which would lead to the amplification being applied with a better fidelity. One can easily account for this new ordering of the operations with the commutation relation in Eq.~\eqref{appendix_eq:useful_commutation_relation}. 

We remark that the probabilistic nature of any noiseless amplification protocol would dramatically increases the sampling overhead of our scheme. One way around this involves conducting the amplification offline and storing the successfully amplified state until it is required.

Alternatively, for many experimentally relevant classes of states, one can avoid the burden of actually implementing the amplification step, provided that the protocol's input state is the following amplified version of $\rho_0$, namely
\begin{equation}
    \rho_{\textrm{amp}} = \frac{1}{\mathcal{N}_{\textrm{amp}}(\rho_0)} g_\mu^{\hat{n}} \rho_0 g_\mu^{\hat{n}} \, ,
\end{equation}
where $\mathcal{N}_{\textrm{amp}}(\rho_0) = \Tr[g_\mu^{\hat{n}} \rho_0 g_\mu^{\hat{n}}]$ is a normalizing factor.
In particular, if the amplified state $\rho_\textrm{amp}$ belongs to the same class of states as $\rho_0$, then noiseless amplification simply corresponds to tuning the parameter that characterize such class.
For example, if $\rho_0$ is a cat state of magnitude $\alpha_0$, then $\rho_{\textrm{amp}}$ is a cat state of magnitude $\alpha_{\textrm{amp}} = g_\mu \alpha_0 > \alpha_0$. 
Similarly, noiselessly amplifying a squeezed vacuum state is equivalent to increasing its squeezing parameter.

In this setting, we may also change perspective and, instead of using $\rho_0$ to define $\rho_{\textrm{amp}} \sim g_\mu^{\hat{n}} \rho_0 g_\mu^{\hat{n}}$, we start from the amplified state $\rho_{\textrm{amp}}$ and use it to determine $\rho_0\sim g_\mu^{-\hat{n}} \rho_{\textrm{amp}} g_\mu^{-\hat{n}} $, i.e., the state whose noise-free expectation value we are estimating.
This approach is particularly helpful when experimental limitations constrain the range of easily accessible states, as illustrated in the following example.
Suppose that the most energetic coherent state we have experimental access to is $\ket{\alpha_{\textrm{max}}}$. It follows that, using $\rho_{\textrm{amp}} = \ketbra{\alpha_{\textrm{max}}}$ as an initial state, we can estimate noise-free expectation values associated with the state $\ket{{\alpha_{\textrm{max}}}/{g_\mu}}$.


\subsubsection{\label{subsubsec:photon_subtraction}Photon Subtraction}
Photon subtraction can be probabilistically implemented using a beam splitter and photon number resolving detector \cite{kim2008recent}. 
Consider a beam splitter of transmissivity $1 - \mu$, its unitary operation defined by $U_{\textrm{BS},\mu} = e^{\arccos(\sqrt{1 - \mu})(\hat{a}_A^\dag \hat{a} - \hat{a}^\dag \hat{a}_A)}$, where the $A$ subscript denotes the ancillary mode. 
If $\rho_{\textrm{amp}}$ is sent through the beam splitter together with a vacuum state in the second input port, then a measurement of $j$ photons in the output port of the ancillary mode will herald the state (details can be found in
Appendix~\ref{appendix_sec:photon_subtraction})
\begin{equation}
    \rho_{\textrm{amp}} \rightarrow \frac{1}{p_j} K_j (\mu) \rho_{\textrm{amp}} K_j^{\dag} (\mu) \, ,
    \label{eq:heralded_state}
\end{equation}
where $K_j (\mu)$ are the Kraus operators defined in Eq.~\eqref{eq:Kraus_operators}, and $p_j=\Tr[K_j (\mu) \rho_{\textrm{amp}} K_j^\dag (\mu)]$ is the probability of having measured $j$ photons.
Finally, comparing Eq.~\eqref{eq:heralded_state} with Eq.~\eqref{eq:noisy_state_j} reveals that, for the heralded state to be equal to $\mathcal{E}_j [\rho_0]$, the amplification factor must be set to $g_\mu = 1 / \sqrt{(1 - \gamma)(1 - \mu)}$.
Hence, assuming that we start our protocol with the amplified state $\rho_{\textrm{amp}}$, the probabilistic nature of the photon subtraction operation naturally allows us to sample from the set of states $\{\mathcal{E}_j [\rho_0]\}$ according to the probability distribution $\{p_j\}$. This is depicted graphically in Fig.~\ref{fig:single_mode_channel_selection/full_circuit_diagram}.


\subsubsection{\label{subsubsec:post-processing}Post-Processing}
Since the implementation of the maps $\lbrace\mathcal{E}_j\rbrace$ is inherently probabilistic, Monte-Carlo methods can not be straightforwardly used to estimate $\expval{O}_{\textrm{ideal}}$.
Nevertheless, the linearity of the expectation value lets us write
\begin{equation}
    \expval{O}_{\textrm{ideal}} = \sum_{j = 0}^\infty \omega_j(\rho_0) \expval{O}_{j, \textrm{noisy}} \, ,
\end{equation}
where $\expval{O}_{j, \textrm{noisy}} = \Tr[O \, (\mathcal{U}_{\textrm{noisy}} \circ \mathcal{E}_j) [\rho_0] ]$ is the noisy expectation value of $O$ given that the $j^{\textrm{th}}$ channel was implemented. 
In other words, a suitable linear combination of the noisy expectation values $\lbrace\expval{O}_{j, \textrm{noisy}}\rbrace$ allows us to estimate the noise-free expectation value $\expval{O}_{\textrm{ideal}}$.
In $N$ runs of the circuit depicted in Fig.~\ref{fig:single_mode_channel_selection/full_circuit_diagram}, we collect $N$ measurement outcomes $\lbrace O^{(j_\ell)}_\ell\rbrace_{\ell=1}^N$, where the superscript $j_\ell$ denotes how many photons were measured by the photon number resolving detector (i.e., which particular channel $\mathcal{E}_{j_\ell}$ was implemented) on the $\ell^{\textrm{th}}$ run.
The noisy expectation value $\expval{O}_{k, \textrm{noisy}}$ is estimated via the sample average
\begin{equation}
    \bar{O}_{k, \textrm{noisy}} = \frac{1}{N_k} \sum_{\ell|j_\ell = k} O_\ell^{(j_\ell)} \, ,
\end{equation}
where $N_k$ is the number of times $\mathcal{E}_k$ was implemented. 
Notice that the probability of subtracting $j + 1$ photons (if possible) is typically a factor of $\mathcal{O}(\frac{\mu}{(j + 1)(1-\mu)})$ smaller than subtracting $j$ photons. We will therefore assume to have a set of noisy expectation values of increasing $j$ up to some maximum cutoff, $J_{\textrm{max}}$, i.e., we have $\lbrace \bar{O}_{j, \textrm{noisy}} \rbrace_{j = 0}^{J_{\textrm{max}}}$. We then linearly combine these expectations values to obtain the mitigated estimator of $\expval{O}_{\textrm{ideal}}$, namely
\begin{equation}
    \bar{O}_{\textrm{mit}} = \sum_{j = 0}^{J_{\textrm{max}}} \omega_j (\rho_0) \bar{O}_{j, \textrm{noisy}} \, ,
    \label{eq:mitigated_estimator}
\end{equation}
which has an expected value of 
\begin{equation}
    \mathbb{E} [\bar{O}_{\textrm{mit}}] = \sum_{j = 0}^{J_{\textrm{max}}} \omega_j (\rho_0) \expval{O}_{j, \textrm{noisy}} \, .
\end{equation}
We may formally express the previous equation as $\mathbb{E} [\bar{O}_{\textrm{mit}}] = \Tr[O \rho_{\textrm{mit}}]$, where 
\begin{equation}
\label{eq:pseudo_state_definition}
    \rho_{\textrm{mit}} = \sum_{j = 0}^{J_{\textrm{max}}} \omega_j (\rho_0) (\mathcal{U}_{\textrm{noisy}} \circ \mathcal{E}_j) [\rho_0] 
\end{equation}
is a \emph{pseudo-state}, as it may not be a valid density matrix. 
Nevertheless, the fidelity between $\rho_{\textrm{mit}}$ and $\mathcal{U}_{\textrm{ideal}} [\rho_0]$ quantifies the ability of our error mitigation protocol to recover the noise-free expectation value for generic observables \cite{cai2021practical}.


\subsection{\label{subsec:error_analysis}Error Analysis}
The bias and variance of the mitigated estimator Eq.~\eqref{eq:mitigated_estimator} for a particular choice of $J_{\textrm{max}}$ are respectively given by
\begin{align}\label{eq:bias_and_variance_of_mitigated_expectation_value_estimator}
    \textrm{Bias} [\bar{O}_{\textrm{mit}}; J_{\textrm{max}}] = \left| \sum_{j = J_{\textrm{max}} + 1}^\infty \omega_j (\rho_0) \expval{O}_{j, \textrm{noisy}}\right|, \\
    \textrm{Var} [\bar{O}_{\textrm{mit}}; J_{\textrm{max}}] = \sum_{j = 0}^{J_{\textrm{max}}} \frac{|\omega_j (\rho_0)|^2}{N_j} \textrm{Var} [O_{j, \textrm{noisy}}] \, ,
\end{align}
where $\textrm{Bias}[\bar{O}_{\textrm{mit}}; J_{\textrm{max}}] := | \mathbb{E}[\bar{O}_{\textrm{mit}|J_{\textrm{max}}}] - \expval{O}_{\textrm{ideal}}|$ and $\textrm{Var} [O_{j, \textrm{noisy}}] = \expval{O^2}_{j, \textrm{noisy}} - \expval{O}_{j, \textrm{noisy}}^2$.

We point out that both the bias and variance fundamentally arise from the finite number of samples considered and therefore vanish in the infinite sampling limit. For a fixed (and suitably large) choice of $N$, the value of $J_{\textrm{max}}$ can be increased by increasing $\mu$. As $J_{\textrm{max}}$ increases, the bias tends to decrease while the variance (and hence sampling overhead) monotonically increases - this accuracy/overhead trade-off is characteristic of many QEM schemes.

In practice, $J_{\textrm{max}}$ is implicitly defined as the smallest integer $J$ that satisfies $\textrm{Bias} [\bar{O}_{\textrm{mit}}; J] < \vartheta_{\textrm{bias}}$, where $\vartheta_{\textrm{bias}}$ is the chosen acceptable level of bias that we want to achieve. As the channel $\mathcal{E}_{J_{\textrm{max}}}$ is implemented with probability $p_{J_\textrm{max}}$, it follows that the number of samples required to achieve $\vartheta_{\textrm{bias}}$ scales as $\mathcal{O} (p_{J_{\textrm{max}}}^{-1})$. 
Once $J_\textrm{max}$ is fixed, any measurement result corresponding to the $k^{\textrm{th}}$ channel for $k > J_{\textrm{max}}$ should be discarded because these results will unnecessarily increase the shot-noise.

It is clear that in experimental settings Eq.~\eqref{eq:bias_and_variance_of_mitigated_expectation_value_estimator} can not be used to estimate the bias, since its expression explicitly depends on unknown noisy expectation values.
Nonetheless, for non-zero $\expval{O}_{\textrm{ideal}}$ the \emph{fractional} bias (or equivalently percentage bias), $\textrm{Bias} [\bar{O}_{\textrm{mit}}; J] / \expval{O}_{\textrm{ideal}}$, may still be approximately estimated if we can assume that $\expval{O}_{j, \textrm{noisy}} / \expval{O}_{\textrm{ideal}} \approx \mathcal{O} (1)$. With this simplification, the fractional bias reads
\begin{equation}
    \frac{\textrm{Bias} [\bar{O}_{\textrm{mit}}; J]}{\expval{O}_{\textrm{ideal}}} \approx \left|\sum_{j = J_{\textrm{max}} + 1}^\infty  \omega_j (\rho_0) \right| \, ,
\end{equation}
and we can therefore estimate the $J_{\textrm{max}}$ necessary to achieve a target bias using the coefficients $\{\omega_j (\rho_0)\}$ only.
Finally, assuming $N_j \approx p_j N$, the sampling overhead reads
\begin{equation}
    \label{eq:sampling_overhead_single_mode}
    \textrm{sampling overhead } \approx \sum_{j = 0}^{J_{\textrm{max}}} \frac{|\omega_j (\rho_0)|^2}{p_j} \frac{\textrm{Var} [O_{j, \textrm{noisy}}]}{\textrm{Var} [O_{\textrm{noisy}}]}.
\end{equation}
This can also be estimated using only $\{ \omega_j \}$ and $\{p_j\}$ under the assumption that $\textrm{Var} [O_{j, \textrm{noisy}}] / \textrm{Var} [O_{\textrm{noisy}}] \approx \mathcal{O}(1)$.

\subsection{\label{subsec:monte-carlo_approach}Monte-Carlo Approach}
For specific classes of input states $\rho_0$ it may be possible to efficiently prepare each state $\mathcal{E}_j[\rho_0]$ appearing in the inverse noise channel decomposition Eq.~\eqref{eq:inverse_photon_loss_decomposition_physcially_realizable_channels}, without having to directly implement the amplification and the photon subtraction (probabilistic) operations.
In this scenario, our protocol can be implemented using the standard Monte-Carlo technique. Explicitly, in each experimental shot one selects an initial state from the ensemble $\{\mathcal{E}_j[\rho_0]\}$ with probability $\{q_j = |\omega_j| / S\}$ where $S = \sum_{j = 0}^\infty |\omega_j|$. The measurement outcome in each shot is then multiplied $S \textrm{sgn}(\omega_j)$ before being averaged over.
This provides an unbiased estimate of $\expval{O}_{\textrm{ideal}}$ because there is no finite cutoff acting on the operator-sum representation of the inverse loss channel in Eq.~\eqref{eq:inverse_photon_loss_decomposition_physcially_realizable_channels}. This Monte-Carlo approach can then be easily generalised to the multi-mode setting (see Sec.~\ref{appendix_subsec:monte-carlo_approach_multiple-modes} for details).

In Sec.~\ref{subsec:cat_states}, we provide a detailed example of the Monte-Carlo approach for the two-component cat state - for these states, Eq.~\eqref{eq:inverse_photon_loss_decomposition_physcially_realizable_channels} can be rewritten in such a way that it contains only two terms and hence we just have to sample from two different initial states, $\{\mathcal{E}_0 [\rho_0], \mathcal{E}_1 [\rho_0]\}$.

Another notable example is given by the single-photon Fock state $\rho_0 = \ketbra{1}$.
In this case, one easily shows that $\Lambda_{\gamma}^{-1}[\rho_0] = \frac{1}{1 - \gamma}\ketbra{1} - \frac{\gamma}{1 - \gamma} \ketbra{0}$. 
Hence, one can implement error cancellation by sending into the noisy circuit $\ket{1}$ ($\ket{0}$) with probability $1 / (1 - \gamma)$ ($\gamma / (1 - \gamma)$) and applying a post-processing weight of $S=\frac{1 + \gamma}{1 - \gamma}$ ($-S$) to the measurement outcome.


\subsection{\label{sec:examples}Examples}
In this section, we provide explicit examples of our error cancellation protocol to showcase its capabilities and assess its performance.
In particular, we will consider the following classes of input states: single-mode squeezed vacuum states and two-component cat states.


\subsubsection{\label{subsec:squeezed_states}Squeezed States}
We first consider the input state to be a single mode squeezed vacuum state, i.e., $\rho_0 = \ketbra{S(r_0)}$, defined as
\begin{align}
\begin{split}
\label{eq:single_mode_squeezed_vacuum_state_SMSV_definition}
    \ket{S(r_0)} &=  S(r_0) \ket{0} \\
    &=  \frac{1}{\sqrt{\cosh(r_0)}}\sum_{n = 0}^\infty \left(\frac{\tanh(r)}{2} \right)^n \frac{\sqrt{(2n)!}}{n!} \ket{2n} \, .
\end{split}
\end{align}
Here, $r_0 \geq 0$ is the squeezing parameter and $S(r_0) = e^{\frac{r_0}{2}(\hat{a}^2 - \hat{a}^{\dag 2} )}$ is the single mode squeezing operator.
As previously mentioned, noiselessly amplifying $\ket{S(r_0)}$ by a factor of $g>1$ produces a new squeezed vacuum state $\ket{S(r_{\textrm{amp}})} \propto g^{\hat{n}}S(r_0)\ket{0}$, where $\tanh(r_{\textrm{amp}}) = g^2 \tanh(r_0)$. Notice that $r_{\textrm{amp}}$ is not well defined when $g^2 \tanh(r_0) \ge 1$. If this is the case then it is not physically meaningful to exactly amplify $\ket{S(r_0)}$ by a factor $g$ \footnote{Since $g^{\hat{n}}$ is an unbounded operator, we can not expect it to produce physical states when acting on arbitrary CV states. This is a generic property of unbounded operators acting on states in an infinite dimensional Hilbert space.}. Therefore, in order to exactly amplify by increasing the squeezing of the initial state, one has to either decrease $\mu$ or redefine $r_0$ to $r_0 \rightarrow \tilde{r_0} < r_0$, both of which come with potential consequences.
Alternatively, one could amplify by $g$ within some subspace, as discussed in Sec.~\ref{subsubsec:amplification}. For example, in Ref.~\cite{PhysRevLett.128.160501_teleamplification}, the authors show that their noiseless linear teleamplification scheme is able to approximately amplify single mode squeezed vacua beyond this $g^2 \tanh(r_0) < 1$ limit.

To assess the performance of our mitigation protocol, we consider the scenario where the ideal unitary evolution $\mathcal{U}_{\textrm{ideal}}$ is the identity and the measured observable corresponds to a projector onto the (pure) noiseless reference state, $\rho_0$, i.e., we consider the observable $O = \rho_0$.
Clearly, the expectation value of $\rho_0$ with respect to an arbitrary state $\sigma$ corresponds to their quantum fidelity, namely $\Tr[\rho_0 \sigma] = F(\rho_0,\sigma)$ \cite{cai2021practical, nielsen2010quantum}. Hence, in this scenario, the mitigated expectation value corresponds to the fidelity $F(\rho_0, \rho_{\textrm{mit}})$ between the target state $\rho_0$ and the pseudo-state $\rho_{\textrm{mit}}$ defined in Eq.~\eqref{eq:pseudo_state_definition}. 
Intuitively, we expect that if the state $\sigma$ has a large fidelity with $\rho_0$, then $\Tr[A \, \sigma]$ provides a good estimate of $\Tr[A \, \rho_0]$ for most observables $A$. A value of $F(\rho_0, \rho_{\textrm{mit}})$ close to one then implies that (for most choices of $A$ and generic unitary dynamics $\mathcal{U}$) $\Tr[A \, \mathcal{U} [\rho_{\textrm{mit}}]]$ will be a good estimator of $\Tr[A \, \mathcal{U} [\rho_0]]$.
We therefore can use $F(\rho_0, \rho_{\textrm{mit}})$ as a metric of the mitigation scheme's performance, independent of any particular dynamics or observables.

\begin{figure}
    \centering
    \includegraphics[width=0.48\textwidth]{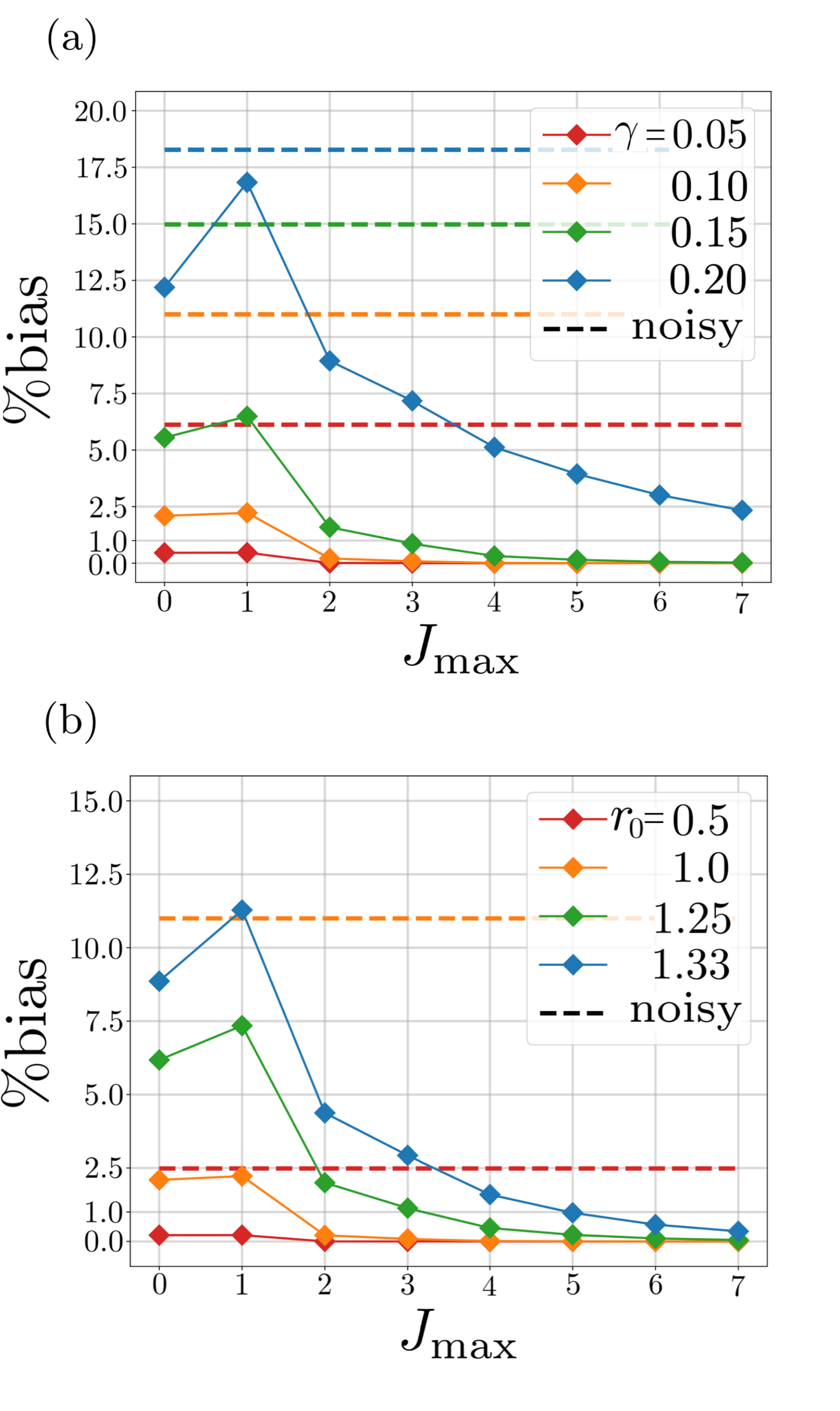}
    \caption{Percentage bias as a function of $J_{\textrm{max}}$ for different values of $r_0$ and $\gamma$, for the estimation of the fidelity with respect to the noiseless target state.
    The dashed horizontal lines represent the percentage bias of the \emph{unmitigated} estimators. 
    In (a), we vary $\gamma$ at fixed $r_0 = 1$.
    On the other hand, in (b) we fix $\gamma = 0.1$ and consider different values of $r_0$.
    The unmitigated percentage biases for $r_0 = 1$ and $r_0 = 1.33$ lie outside the graph, but are respectively given by $18.0\%$ and $20.6\%$.
    In all displayed cases, we see an improvement over the unmitigated result, even for modest values of $J_{\textrm{max}}$.}
    \label{fig:bias_SMSV}
\end{figure}

In Fig.~\ref{fig:bias_SMSV}, we plot the exact percentage bias of $\bar{O}_{\textrm{mit}}$ with $O = \ketbra{S(r_0)}$ as a function of $J_{\textrm{max}}$, for different values of loss, $\gamma$, and squeezing, $r_0$.
We see significant improvements with respect to the unmitigated results (horizontal lines), indicating that the pseudo-state $\rho_{\textrm{mit}}$ has a higher fidelity with $\rho_0$ than $\Lambda_\gamma [\rho_0]$. For example, to ensure $\vartheta_{\textrm{bias}} < 10^{-3}$ (corresponding to a percentage bias smaller than $0.1\%$) with $\gamma = 0.1$ and $r_0 = 1.0$ only requires $J_{\textrm{max}} = 3$.
Even when $J_{\textrm{max}} = 0$, which corresponds to performing a zero photon subtraction \cite{zero_photon_subtraction} on the amplified state, we see an improvement in the bias for all plotted cases.

\begin{figure}
    \centering
    \includegraphics[width = 0.48\textwidth]{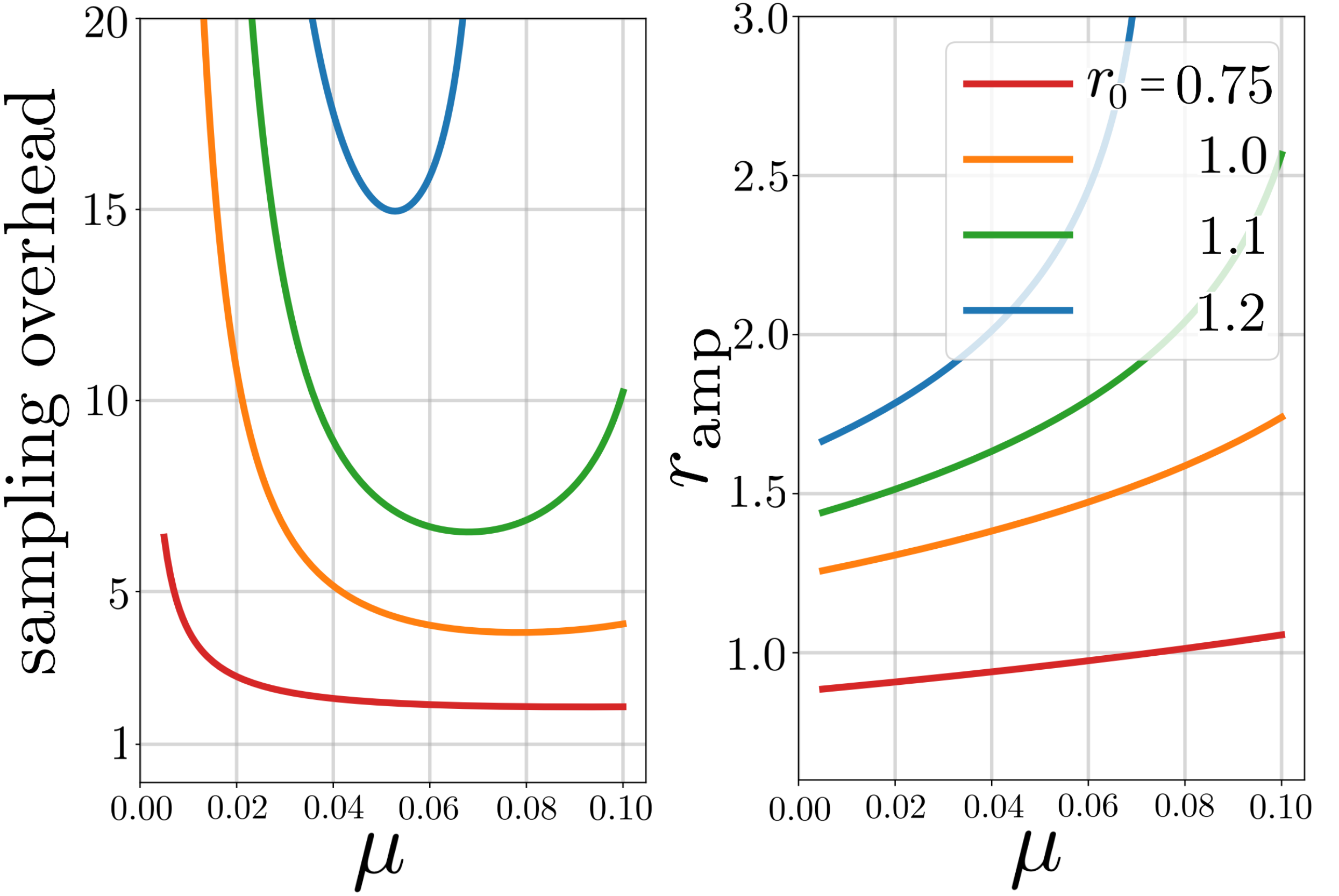}
    \caption{Sampling overhead and $r_{\textrm{amp}}$ as a function of $\mu$ for varied $r_0$ with fixed $\gamma = 0.1$.
    For $r_0 = (0.75, 1.0, 1.1, 1.2)$, we choose $J_{\textrm{max}} = (1, 2, 3, 3)$ which gives the percentage biases as $(0.72\%, 0.21\%, 0.23\%, 0.65\%) < 1\%$. The corresponding \emph{unmitigated} biases are $(5.86\%, 11.0\%, 13.58\%, 16.46\%)$. We observe a trade-off between minimising the sampling overhead and minimising the required amount of squeezing. The optimal choice of $\mu$ will depend the individual experimental details such as the difficulty in preparing highly squeezed states and circuit shot rate. We point out that we did not include the effect of discarding detection outcomes corresponding to $J>J_{\textrm{max}}$ in the computation of the sampling overhead and hence we are slightly underestimating its true value.
    Fixing $r_0$ and varying $\gamma$ produces analogous looking behaviour.
    }
    \label{fig:sampling_overhead_squeezing_against_mu}
\end{figure}

\begin{figure}
    \centering
    \includegraphics[width = 0.48\textwidth]{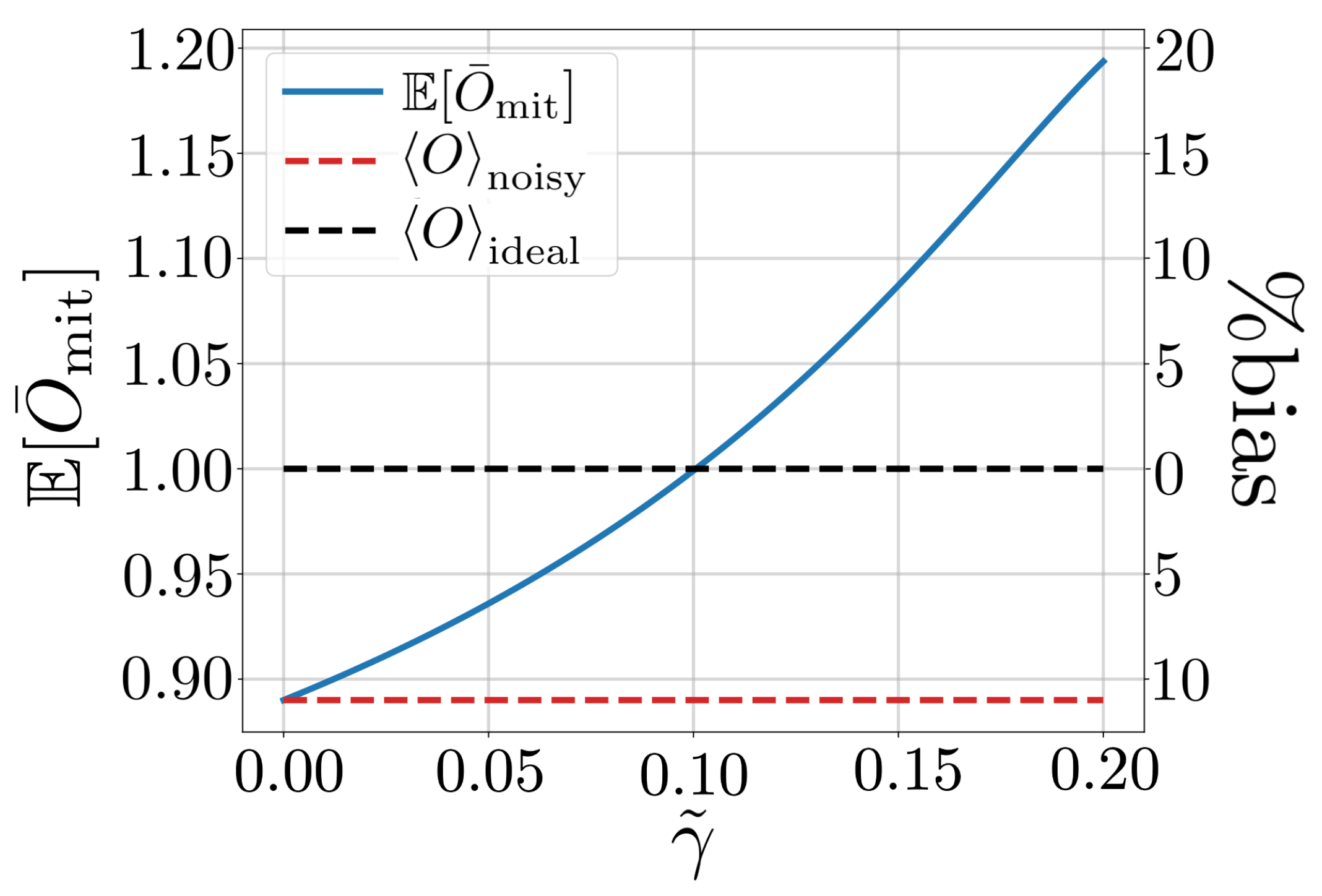}
    \caption{$\mathbb{E} [\bar{O}_{\textrm{mit}}] = F(\rho_0, \rho_{\textrm{mit}})$ as a function of $\tilde{\gamma}$, with $\gamma = 0.1$, $r_0 = 1$ and $J_{\textrm{max}} = 3$.
    These parameters gives a mitigated percentage bias of $0.1\%$ when we have perfect knowledge of the loss parameter, i.e., $\tilde{\gamma} = \gamma$.
    We observe that our mitigation scheme is stable under imperfect knowledge of $\gamma$: for example, $\tilde{\gamma}=0.075$ (a $25\%$ error on the noise parameter characterization) only leads to a modest $4\%$ bias of the mitigated estimator.}
    \label{fig:imperfect_gamma_knowledge}
\end{figure}

In Fig.~\ref{fig:sampling_overhead_squeezing_against_mu}, we plot the sampling overhead and $r_{\textrm{amp}}$ as a function of $\mu$ for different values of $r_0$ and fixed $\gamma$. We choose the acceptable level of bias, $\vartheta_{\textrm{bias}}$, to be $1\%$.
For suitably small $\mu$, we observe that increasing $\mu$ tends to decrease the sampling overhead. This is because it increases the probability that $j \ne 0$ photons are subtracted which in turn will tend to decrease the shot noise in $\bar{O}_{j \ne 0, \textrm{noisy}}$. This comes at the cost of increasing the necessary level of squeezing, $r_{\textrm{amp}}$. 
Finding the optimal value of $\mu$ in the trade-off between sampling overhead and $r_{\textrm{amp}}$ will depend on experimental considerations such as the difficulty of increasing the squeezing parameter and the circuit shot rate.
\\
\\
Finally, we investigate the stability of our error cancellation protocol in the presence of imperfect noise strength parameter estimation. In particular, we imagine that the experimenter has incorrectly estimated the loss parameter to be $\tilde{\gamma}$ and then study the effects that implementing the mitigation scheme using loss parameter $\tilde{\gamma} \ne \gamma$ has on the bias.
In Fig.~\ref{fig:imperfect_gamma_knowledge}, we plot $\mathbb{E}[\bar{O}_{\textrm{mit}}] = F(\rho_0, \rho_{\textrm{mit}})$ as a function of $\tilde{\gamma}$ with $r_0 = 1$, $\gamma = 0.1$ and $J_{\textrm{max}} = 3$ - for this this choice of parameters, when $\tilde{\gamma} = \gamma$ the percentage bias is $< 0.1\%$.

We see that the percentage bias then grows approximately linearly as $\tilde{\gamma}$ moves away from the true loss parameter, and that the gradient of this curve is favourable - for example, a $25\%$ error in estimation of the noise parameter leads to only a $4\%$ bias in the expected value of the mitigated estimator. In order for the mitigated bias to match the noisy expectation value bias (which for these parameters is around $11\%$), one would have to estimate $\tilde{\gamma} = 0$ or $\tilde{\gamma} = 0.16$. Therefore, for these parameters as long as the experimental error in $\gamma$ estimation is below $60\%$, the bias of the mitigated estimator outperforms the noisy results. Notably, for all $\tilde{\gamma} < \gamma$ our mitigation scheme outperforms the noisy result and therefore even partially undoing the losses is always better (from a bias-perspective) than running the noisy dynamics.

This slow growth of the bias with respect to $|\gamma - \tilde{\gamma}| / \gamma$ constitutes a strong indication that our mitigation scheme is robust against imperfect characterisation of the noise strength. This is because the impact of estimating $\tilde{\gamma} \ne \gamma$ presents itself as a moderate bias in the end result, but this bias is relatively small for reasonable estimates of $\gamma$, such as within $10\%$.

\subsubsection{\label{subsec:cat_states}Cat States}
\begin{figure}
    \centering
    \includegraphics[width=0.48\textwidth]{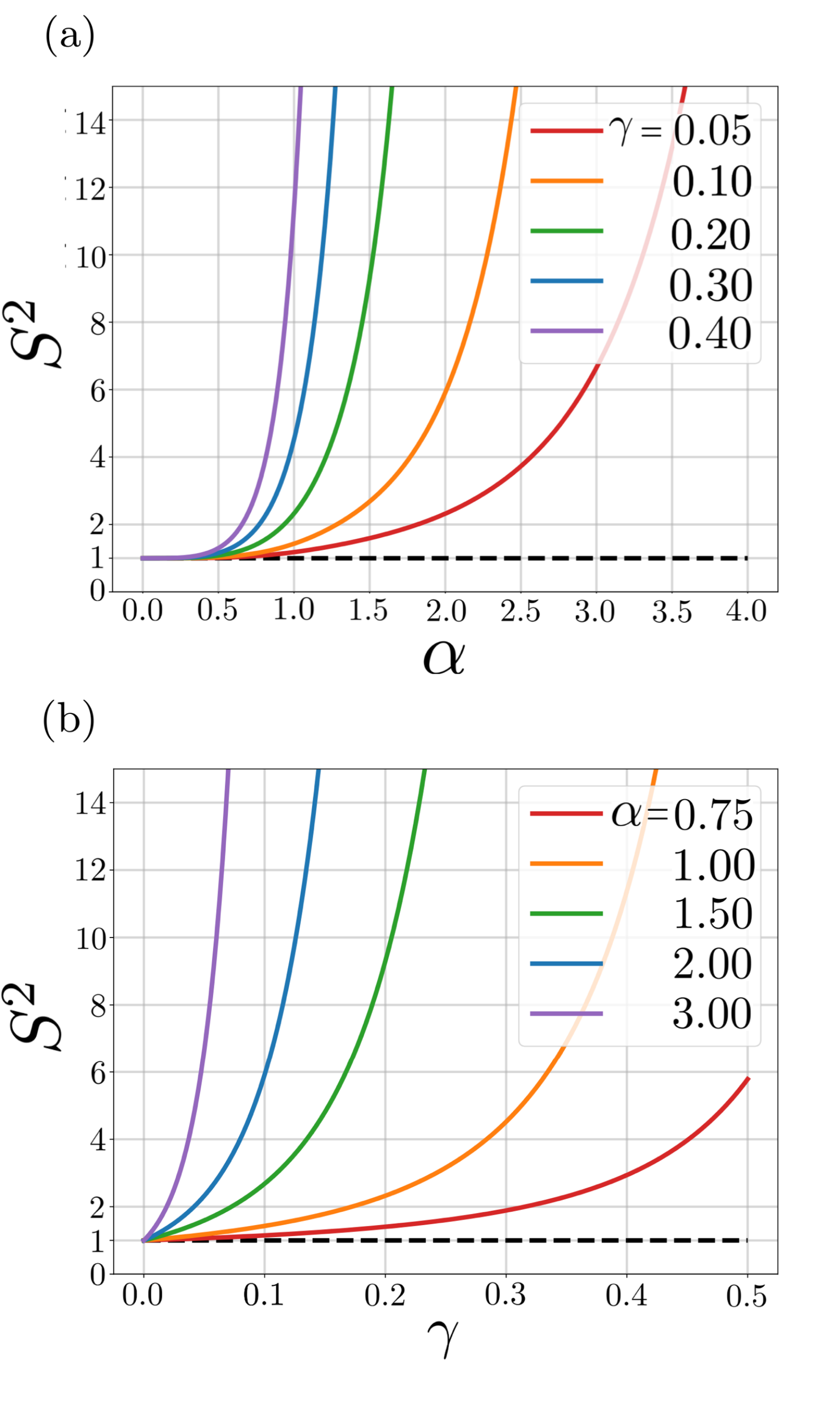}
    \caption{$S^2 \approx$ sampling overhead as a function of (a) $\gamma$ and (b) $|\alpha|$. When $\textrm{Var} [O_{\textrm{even}}] \approx \textrm{Var} [O_{\textrm{odd}}] \approx \textrm{Var} [O_{\textrm{noisy}}]$, we can approximate the sampling overhead with $S^2$, which is plotted. Despite the exponential growth with respect to $|\alpha|$ and $\gamma$, there is a large and relevant parameter space in which the sampling overhead is sufficiently modest to be experimentally reasonable.}
    \label{fig:cat_state_sampling_overhead_against_alpha_and_gamma}
\end{figure}

In this section, we provide an example of the Monte-Carlo approach previously outlined in Section ~\ref{subsec:monte-carlo_approach}.
We do so by considering two-component cat states  $\rho_\phi(\alpha) = \ketbra{\mathcal{C}_\phi (\alpha)}$ as input states for our mitigation scheme.
These are defined as $\ket{\mathcal{C}_{\phi}(\alpha)} = (\ket{\alpha} + e^{i \phi} \ket{-\alpha}) / \mathcal{A}_\phi(\alpha)$, where $\mathcal{A}_\phi(\alpha) = \sqrt{2(1 + \cos(\phi)e^{-2|\alpha|^2})}$ denotes the normalisation coefficient.
Since the two-component cat states are eigenstates of $\hat{a}^2$, it follows that we only need the channels $\mathcal{E}_0$ and $\mathcal{E}_1$ to fully describe the action of the inverse noise channel $\Lambda_\gamma^{-1}$. One can easily show that
\begin{align}
\begin{split}
    \mathcal{E}_{j} [\rho_\phi (\alpha)] = \begin{cases}
        \mathcal{E}_0 [\rho_\phi (\alpha)] & j \textrm{ even}\\
        \mathcal{E}_1 [\rho_\phi (\alpha)] & j \textrm{ odd}
    \end{cases}
\end{split}
\end{align}
which further implies that we can write
\begin{equation}
    \Lambda_\gamma^{-1} [\rho_\phi (\alpha)] = W_{\textrm{even}} \rho_\phi (g_0 \alpha) + W_{\textrm{odd}} \rho_{\phi + \pi} (g_0 \alpha) \, ,
\end{equation}
where the weights are given by
\begin{align}
\begin{split}
    W_{\textrm{even}} & = \sum_{j=0}^\infty \omega_{2j}=  e^{\frac{\gamma |\alpha|^2}{1 - \gamma}} \textrm{cosh} \left(\frac{\gamma |\alpha|^2}{1 - \gamma}\right) \left| \frac{\mathcal{A}_\phi (g_0 \alpha)}{\mathcal{A}_\phi (\alpha)} \right|^2 \, ,  \\
    W_{\textrm{odd}} & =  \sum_{j=0}^\infty \omega_{2j+1}= - e^{\frac{\gamma |\alpha|^2}{1 - \gamma}} \textrm{sinh} \left(\frac{\gamma |\alpha|^2}{1 - \gamma}\right) \left| \frac{\mathcal{A}_{\phi + \pi} (g_0 \alpha)}{\mathcal{A}_\phi (\alpha)} \right|^2 \, . 
\end{split}
\end{align}
We remind the reader that $g_0 = 1 / \sqrt{1 - \gamma}$ and point out that $W_{\textrm{odd}}$ is negative.
If one is able to directly prepare the states $\ket{\mathcal{C}_{\phi} (g_0 \alpha)}$ and $\ket{\mathcal{C}_{\phi + \pi} (g_0 \alpha)}$, then the mitigation scheme may be implemented simply by sending these two states through the circuit with probabilities respectively given by $q_{\textrm{even}} = |W_{\textrm{even}}|/S$ and $q_{\textrm{odd}} = |W_{\textrm{odd}}|/S$, where
\begin{equation}
    S = |W_{\textrm{even}}| + |W_{\textrm{odd}}| = \frac{e^{\frac{2\gamma}{1 - \gamma} |\alpha|^2} + \textrm{cos} (\phi) e^{-\frac{2}{1 - \gamma} |\alpha|^2}}{1 + \textrm{cos} (\phi) e^{-2|\alpha|^2}}.
    \label{eq:S}
\end{equation}
The mitigated estimator is then given by
\begin{equation}
    \bar{O}_{\textrm{mit}} = S \left( \bar{O}_{\textrm{even}} - \bar{O}_{\textrm{odd}} \right).
\end{equation}
The latter is unbiased because we did not include a cutoff in the operator-sum representation of $\Lambda_{\gamma}^{-1}$.
The associated sampling overhead reads
\begin{align}
\begin{split}
    \textrm{sampling overhead} &\approx S \, \bigg( |W_{\textrm{even}}| \frac{\textrm{Var} [O_{\textrm{even}}
    ]}{\textrm{Var} [O_{\textrm{noisy}}]} \\
    &\,\,\,\,\,\, + |W_{\textrm{odd}}| \frac{\textrm{Var} [O_{\textrm{odd}}]}{\textrm{Var} [O_{\textrm{noisy}}]} \bigg).
\end{split}
\end{align}
When $\textrm{Var} [O_{\textrm{even}}] \approx \textrm{Var} [O_{\textrm{odd}}] \approx \textrm{Var} [O_{\textrm{noisy}}]$, the sampling overhead is approximately $S^2$, exactly as in the DV PEC case. 
This scaling arises because the task has been reduced to a Monte-Carlo estimation task.
Notice that the leading term in  Eq.~\eqref{eq:S} is exponential in $\gamma |\alpha|^2 / (1 - \gamma)$.
In Figure \ref{fig:cat_state_sampling_overhead_against_alpha_and_gamma} we plot $S^2$ (i.e., the approximate sampling overhead) as a function of $\alpha$ and $\gamma$.

\section{\label{sec:multiple_modes}Extension to Multiple Modes}
In this section, we extend our error cancellation protocol to $M$ modes. Conceptually, the multi-mode protocol is exactly analogous to the single mode case. In order to highlight this similarity, we introduce two new operators corresponding to a multi-mode photon subtraction and a multi-mode amplification/attenuation:
\begin{align}
    \hat{\mathcal{A}}_{\bm{j}}^{(M)} := \bigotimes_{i = 1}^M \hat{a}_i^{j_i} \, ,
\qquad
    \hat{\mathcal{G}}_{\bm{g}}^{(M)} := \bigotimes_{i = 1}^M g_i^{\hat{n}_i}.
\end{align}
Here, $\bm{j} = (j_1, \dots, j_M)$ is a vector of photon subtractions and $\bm{g} = (g_1, \dots, g_M)$ is a vector of the amplification factors $g_i$ in the $i^{\textrm{th}}$ mode.

\subsection{\label{multiple_modes_inverse_channel}Inverse Loss Map}
Our proposed mitigation scheme extends naturally from a single mode to multiple modes when the photon losses are of the form presented in Eq.~\eqref{eq_mm_loss_channel}. By comparison with the single-mode case, one can see from the $P$-representation of $\Lambda_{\bm{\gamma}}^{(M)}$ in Eq.~\eqref{mm_loss_channel_P_representation} that the multi-mode inverse loss map is given by a tensor product of $M$ single mode inverse loss maps with loss parameters $\bm{\gamma} = (\gamma_1, \gamma_2, \dots, \gamma_M)$: 
\begin{equation}
    \Lambda_{\bm{\gamma}}^{-1(M)} = \bigotimes_{i = 1}^M \Lambda_{\gamma_i, i}^{-1}.
    \label{multimode_loss_inverse}
\end{equation}
Once again, this map is Hermitian preserving and trace preserving but not completely positive and hence non-physical. Nevertheless, these operators can be expressed as a sum over physically realisable operations with potentially negative coefficients.
We consider the following set of maps $\{\mathcal{E}_{\bm{j}}^{(M)}\}$:
\begin{align}
    \mathcal{E}_{\bm{j}}^{(M)} [\rho] = \frac{1}{\tilde{\mathcal{N}}_{\bm{j}} (\rho)} \hat{\mathcal{A}}_{\bm{j}}^{(M)}
    \hat{\mathcal{G}}_{\bm{g_0}}^{(M)} \, \rho \, \hat{\mathcal{G}}_{\bm{g_0}}^{(M)} \hat{\mathcal{A}}_{\bm{j}}^{(M)\dag},
\end{align}
 where $j_i \in \{0, 1, 2, \dots\}$ denotes how many photons are subtracted from mode $i$,  $\boldsymbol{g_0} = (g_{0,0}, g_{0,1}, \dots, g_{0,M})$ with $g_{0, i} = 1 / \sqrt{1 - \gamma_i}$, and $\tilde{\mathcal{N}}_{\bm{j}} (\rho)= \Tr[\hat{\mathcal{A}}_{\bm{j}}^{(M)} \hat{\mathcal{G}}_{\bm{g_0}}^{(M)} \, \rho \, \hat{\mathcal{G}}_{\bm{g_0}}^{(M)} \hat{\mathcal{A}}_{\bm{j}}^{(M)\dag}]$ is the normalisation. We again refer to these maps as channels for simplicity. 
 If the initial state $\rho$ can be written as a tensor product, $\rho_{\textrm{prod}} = \bigotimes_{i = 1}^M \rho_i$, then the normalisation reduces to $\tilde{\mathcal{N}}_{\bm{j}} (\rho_{\textrm{prod}}) = \prod_{i = 1}^M \mathcal{N}_{j_i} (\rho_i)$ where $\mathcal{N}_{j_i} (\rho_i)$ is defined in Eq.~\eqref{eq:E_j_normalisaion_definition}.

The $M$ mode inverse loss map can then be decomposed into physically realisable channels with potentially negative coefficients via
\begin{equation}
\label{eq:mm_physically_sum_inverse}
    \Lambda_{\bm{\gamma}}^{-1 \, (M)} [\rho] = \sum_{j_1 = 0}^\infty \dots \sum_{j_M = 0}^\infty \Omega_{\bm{j}} (\rho) \mathcal{E}_{\bm{j}}^{(M)} [\rho] \, .
\end{equation}    
Once again, $\Omega_{\bm{j}} = \left( \prod_{i = 1}^M \frac{(-\gamma_i)^{j_i}}{j_i!} \right) \tilde{\mathcal{N}}_{\bm{j}} (\rho)$ is a quasi-probability distribution and hence satisfies $\sum_{j_1, j_2, \dots j_M = 0}^\infty \Omega_{\bm{j}} = 1$. If $\rho$ is a product state then the coefficients reduce to a product of those defined in Eq.~\eqref{eq:omega_coefficient_definition}, i.e., $\Omega_{\bm{j}} (\rho_{\textrm{prod}}) = \prod_{i = 1}^M \omega_{j_i} (\rho_i)$.

\subsection{Quantum Error Cancellation Protocol}

In order to physically implement the set of channels $\{\mathcal{E}_{\bm{j}}^{(M)}\}$, we need to have access to (noiseless) amplification and multi-photon subtractions.
Photon subtractions are routinely implemented via a beam splitter interaction that further attenuates the state.
In order to compensate for this unwanted effect, we need to amplify each mode by a factor $g_{\mu, i} > g_{0, i}$, where $i$ denotes the mode and $g_{\mu, i}$ is a new parameter that depends on the transmissivity of the beam splitter in the $i^{\textrm{th}}$ mode.
We then redefine the vector of amplification factors as $\bm{g_{\mu}} = (g_{\mu, 1}, g_{\mu, 2}, \dots, g_{\mu, M})$. 
For an ideal initial state $\rho_0$, the amplified initial state, $\rho_{\textrm{amp}}$, is given by
\begin{equation}
    \rho_{\textrm{amp}} = \frac{1}{\tilde{\mathcal{N}}_{\textrm{amp}}} \hat{\mathcal{G}}_{\bm{g_{\mu}}}^{(M)} \rho_0 \hat{\mathcal{G}}_{\bm{g_\mu}}^{(M)},
\end{equation}
where $\tilde{\mathcal{N}}_{\textrm{amp}}= \Tr[\hat{\mathcal{G}}_{\bm{g_{\mu}}}^{(M)} \rho_0 \hat{\mathcal{G}}_{\bm{g_\mu}}^{(M)}]$ is the normalisation. As in the single mode case, we recommend (where possible) placing this amplification in the state preparation stage - the discussion of this amplification within the single mode setting in Sec.~\ref{subsubsec:amplification} is still highly relevant, particularly for product states. For entangled initial states, there are still many experimentally relevant states that maintain their functional form under amplification. For example, amplifying a two mode squeezed vacuum state of squeezing parameter $r_0$ produces a two mode squeezed vacuum of squeezing parameter $r_{\textrm{amp}} > r_0$. Similarly, amplification of two mode entangled coherent state $\sim \ket{\alpha_0}\ket{\beta_0} + \ket{-\alpha_0}\ket{-\beta_0}$ implements the transformations $\alpha_0 \rightarrow \alpha_{\textrm{amp}} > \alpha_0$ and $\beta_0 \rightarrow \beta_{\textrm{amp}} > \beta_0$.
The photon subtraction term can be implemented with $M$ beam splitters and photon number resolving detectors. In each mode, a beam splitter of transmissivity $1 - \mu_i$ for $i \in \{1, 2, \dots, M\}$ is used with a vacuum in the second entry port. The photon number resolving detectors then measure the output of the second exit port. 
A measurement of $\bm{j} = (j_1, j_2, \dots, j_M)$ photons with the photon number resolving detectors will implement a tensor product of Kraus operators,
\begin{equation}
    \mathcal{K}^{(M)}_{\bm{j}, \bm{\mu}} = K_{j_1}(\mu_1) \otimes K_{j_2}(\mu_2) \otimes \dots \otimes K_{j_M}(\mu_M), 
\end{equation}
on $\rho_{\textrm{amp}}$, where $K_{j_i}(\mu_i)$ are defined in Eq.~\eqref{eq:photon_loss_kraus_representation}. This action heralds the state
\begin{align}
\label{appendix_eq:photon_subtraction_output_state}
    \rho_{\textrm{amp}} \rightarrow \frac{1}{\tilde{p}_{\bm{j}}} \mathcal{K}^{(M)}_{\bm{j}, \bm{\mu}} \rho_{\textrm{amp}} \mathcal{K}^{(M) \dag}_{\bm{j}, \bm{\mu}} = \mathcal{E}_{\bm{j}}^{(M)} [\rho_0] \, ,
\end{align}
where $\tilde{p}_{\bm{j}} = \Tr[\mathcal{K}^{(M)}_{\bm{j}, \bm{\mu}} \rho_{\textrm{amp}} \mathcal{K}^{(M) \dag}_{\bm{j}, \bm{\mu}}]$ is the probability that $\bm{j}$ photons were subtracted. The equality in Eq.\eqref{appendix_eq:photon_subtraction_output_state} is easily be seen by expanding both $\rho_{\textrm{amp}}$ and $\mathcal{K}_{\bm{j}, \bm{\mu}}^{(M)}$. We therefore have a way of sampling from $\{\mathcal{E}_{\bm{j}}^{(M)} [\rho_0]\}$ with corresponding probabilities $\{\tilde{p}_{\bm{j}}\}$.

We can now consider measuring a generic $M-$mode observable $O$. It can be seen that the ideal expectation value, $\expval{O}_{\textrm{ideal}} = \Tr[O \, \mathcal{U}_{\textrm{ideal}} [\rho_0] ]$, can be decomposed into
\begin{equation}
\label{appen_eq:ideal_observable_expectation_decomposition}
    \expval{O}_{\textrm{ideal}} = \sum_{\bm{j}} \Omega_{\bm{j}} (\rho_0) \expval{O}_{\bm{j}, \textrm{noisy}} \, ,
\end{equation}
where $\expval{O}_{\bm{j}, \textrm{noisy}} = \Tr[O \, (\mathcal{U}_{\textrm{noisy}} \circ \mathcal{E}_{\bm{j}}^{(M)}) [\rho_0]]$ is the noisy expectation value of $O$ given that the channel $\mathcal{E}_{\bm{j}}$ was implemented and $\sum_{\bm{j}}$ symbolically represents a sum over all possible $\bm{j} = (j_1, j_2, \dots, j_M)$.

Now imagine running the circuit $N$ times. The channel that was implemented in each shot is stochastically determined by the number and distribution of detected photons in the photon subtraction stage. We bin the results based on $\bm{j}$ - the set of different channels implemented is denoted $\bm{J}$. For each element of $\bm{j} \in \bm{J}$, the associated noisy expectation value $\expval{O}_{\bm{j}, \textrm{noisy}}$ is estimated by the arithmetic mean of the $O$ measurement outcomes in which the channel $\mathcal{E}^{(M)}_{\bm{j}}$ was implemented - we denote this $\bar{O}_{\bm{j}, \textrm{noisy}}$. The mitigated estimator of $\expval{O}_{\textrm{ideal}}$ is then
\begin{equation}
    \bar{O}_{\textrm{mit}} = \sum_{\bm{j} \in \bm{J}} \Omega_{\bm{j}} (\rho_0) \bar{O}_{\bm{j}, \textrm{noisy}} \, .
\end{equation}

\subsection{Error Analysis}
The bias and variance of the mitigated estimator $\bar{O}_{\textrm{mit}}$ are given by
\begin{align}
    \textrm{Bias} [\bar{O}_{\textrm{mit}}; \bm{J}] = \left|\sum_{\bm{j} \notin \bm{J}} \Omega_{\bm{j}} (\rho_0) \expval{O}_{\bm{j}, \textrm{noisy}} \right| \, ,
\end{align}
and
\begin{align}
    \textrm{Var} [\bar{O}_{\textrm{mit}}; \bm{J}] = \sum_{\bm{j} \in \bm{J}} \frac{|\Omega_{\bm{j}} (\rho_0)|^2}{N_{\bm{j}}} \textrm{Var} [O_{\bm{j}, \textrm{noisy}}] \, , 
\end{align}
where $\textrm{Var} [O_{\bm{j}, \textrm{noisy}}] = \expval{O^2}_{\bm{j}, \textrm{noisy}} - \expval{O}_{\bm{j}, \textrm{noisy}}^2$ is the single shot variance of $O$ given the channel $\mathcal{E}_{\bm{j}}^{(M)}$ was implemented and $N_{\bm{j}}$ is the number of measurement outcomes associated with $\mathcal{E}_{\bm{j}}^{(M)}$. Once again, the fractional bias can be estimated if one assumes $\expval{O}_{\bm{j}, \textrm{noisy}} / \expval{O}_{\textrm{ideal}} \approx \mathcal{O}(1)$. Given an acceptable bias $\vartheta_{\textrm{bias}}$, a suitable set $\bm{J}$ can be chosen - this choice can have a significant impact on bias. The variance grows monotonically with $|\bm{J}|$ which leads to an accuracy/overhead trade-off that one has to optimize. There also exists an amplification/overhead trade-off governed by $\bm{\mu}$.
Under the approximation $N_{\bm{j}} = \tilde{p}_{\bm{j}} N$, the sampling overhead of the estimated observable $O$ is approximately given by
\begin{equation}
    \textrm{sampling overhead} [O] \approx \sum_{\bm{j} \in \bm{J}} \frac{|\Omega_{\bm{j}} (\rho_0)|^2}{\tilde{p}_{\bm{j}}} \frac{\textrm{Var} [O_{\bm{j}, \textrm{noisy}}]}{\textrm{Var} [O_{ \textrm{noisy}}]} \, , 
\end{equation}
where $\textrm{Var} [O_{\textrm{noisy}}] = \expval{O^2}_{\textrm{noisy}} -\expval{O}_{\textrm{noisy}}^2$ is the single shot variance of $O$ given that only the noisy dynamics, $\mathcal{U}_{\textrm{noisy}}$, were run.
If one is measuring a collection of local observables simultaneously, such as $O_1 := O_1 \otimes \mathbb{I}_2 \otimes \dots \otimes \mathbb{I}_M$, $O_2 := \mathbb{I} \otimes O_2 \otimes \dots \otimes \mathbb{I}_M$, then the sampling overhead is given by the largest individual sampling overhead,
\begin{equation}
    \textrm{sampling overhead} \approx \max_i \left[\textrm{sampling overhead} [O_i] \right] \, .
\end{equation}

\subsection{\label{appendix_subsec:monte-carlo_approach_multiple-modes}Monte-Carlo Approach}
As in the single-mode case, if each \emph{state} in the operator-sum representation of $\Lambda_{\bm{\gamma}}^{-1 (M)}$, namely $\mathcal{E}_{\bm{j}}^{(M)}[\rho_0]$, can be efficiently prepared, then our scheme can be reduced to Monte-Carlo estimation. By sampling from $\{\mathcal{E}_{\bm{j}}^{(M)}[\rho_0]\}$ with probabilities $\{|\Omega_{\bm{j}}| / S\}$ where $S = \sum_{\bm{j}} |\Omega_{\bm{j}}|$, an unbiased estimator of $\expval{O}_{\textrm{ideal}}$ can be obtained by multiplying the measurement outcomes associated with $\bm{j}$ by $S \,  \textrm{sgn}(\Omega_{\bm{j}}(\rho_0))$ - this is exactly analogous to the single-mode Monte-Carlo approach discussed in Sec.~\ref{subsec:monte-carlo_approach}. 

We provide an explicit example of this approach for two-modes in Sec.~\ref{sec:two-mode_examples_multi}.
This approach is also particularly useful for systems in the dual-rail representation \cite{RevModPhys.79.135}, where logical qubits are encoded by a single photon shared between two modes via
\begin{equation}
    \ket{0}_L := \ket{0}\ket{1}, \quad \quad \ket{1}_L := \ket{1}\ket{0}.
\end{equation}
If one wanted to implement our error cancellation scheme using the Monte-Carlo approach for an initial state $\ket{0}_L^{\otimes N} = (\ket{0}\ket{1})^{\otimes N}$ then, under the assumption of uniform losses, $S$ can be shown to be
\begin{equation}
    S = \left(\frac{1 + \gamma}{1 - \gamma}\right)^N.
\end{equation}
Intriguingly, this means one can undo the photon losses by stochastically sending \emph{fewer} single photons into the circuit. We provide some intuition for why this is in Sec.~\ref{subsubsec:fock_intuition_for_loss}.

\subsection{Examples}\label{sec:two-mode_examples_multi}
In this section we look at two classes of entangled initial states, namely two-mode squeezed vacua and entangled coherent states. As for the single-mode squeezed vacuum case, we again use the observable $O = \rho_0$; for the pure initial states considered here, this expectation value estimation task corresponds to estimating the fidelity between $\rho_{\textrm{noisy}}$ and $\rho_{\textrm{mit}}$ with respect to $\rho_0$.
Furthermore, as argued in Sec.~\ref{subsec:squeezed_states}, this provides an observable and unitary-dynamics independent benchmark for our mitigation scheme.
We also consider the covariance matrix as an observable for the two-mode squeezed vacua example.

We allow the loss parameter to fluctuate between shots in Fig.~\ref{fig:covariance_matrix_histogram} and Fig.~\ref{fig:ECS_fidelity_estimate} in order to create a more realistic noise model, as discussed in Sec.~\ref{subsec:noise_model}. Recovery of the ideal expectation value in both of these simulations is strong evidence that our mitigation scheme is robust against a fluctuating loss parameter, and hence demonstrates the applicability of quantum error cancellation realistic scenarios. In Fig.~\ref{fig:ECS_large_varied_loss_histogram}, we plot an archetypal histogram of different loss parameters used in a 1,000,000 shot experiment with average losses exceeding $50\%$.

\subsubsection{Two-Mode Squeezed Vacua}
Consider the two-mode squeezed vacuum (TMSV) states given by
\begin{align}
\begin{split}
    \ket{S_2(r)} &= e^{r (\hat{a}_1^\dag \hat{a}_2^\dag - \hat{a}_1 \hat{a}_2)} \ket{00} \\
    &= \cosh[-1](r) \sum_{n = 0}^\infty \tanh[n](r) \ket{nn}.
\end{split}
\end{align}
Let $\rho_0 = \ketbra{S_2(r_0)}$. Amplifying $\ket{S_2(r_0)}$ by $\mathcal{G}_{\bm{g} = (g_1, g_2)}^{(M)}$ gives $\ket{S_2(r_{\textrm{amp}})}$ up to normalization, where $\tanh(r_{\textrm{amp}}) = g_1 g_2 \tanh(r_0)$ and the amplification normalisation is $\tilde{\mathcal{N}}_{\textrm{amp by } \bm{g}} = \frac{\cosh[2](r_{\textrm{amp}})}{\cosh[2](r_{0})}$. As in the single-mode squeezed vacuum case, we require $g_1 g_2 \tanh(r_0) < 1$ for the amplification to converge and be physically meaningful.

\begin{figure}
    \centering
    \includegraphics[width = 0.49\textwidth]{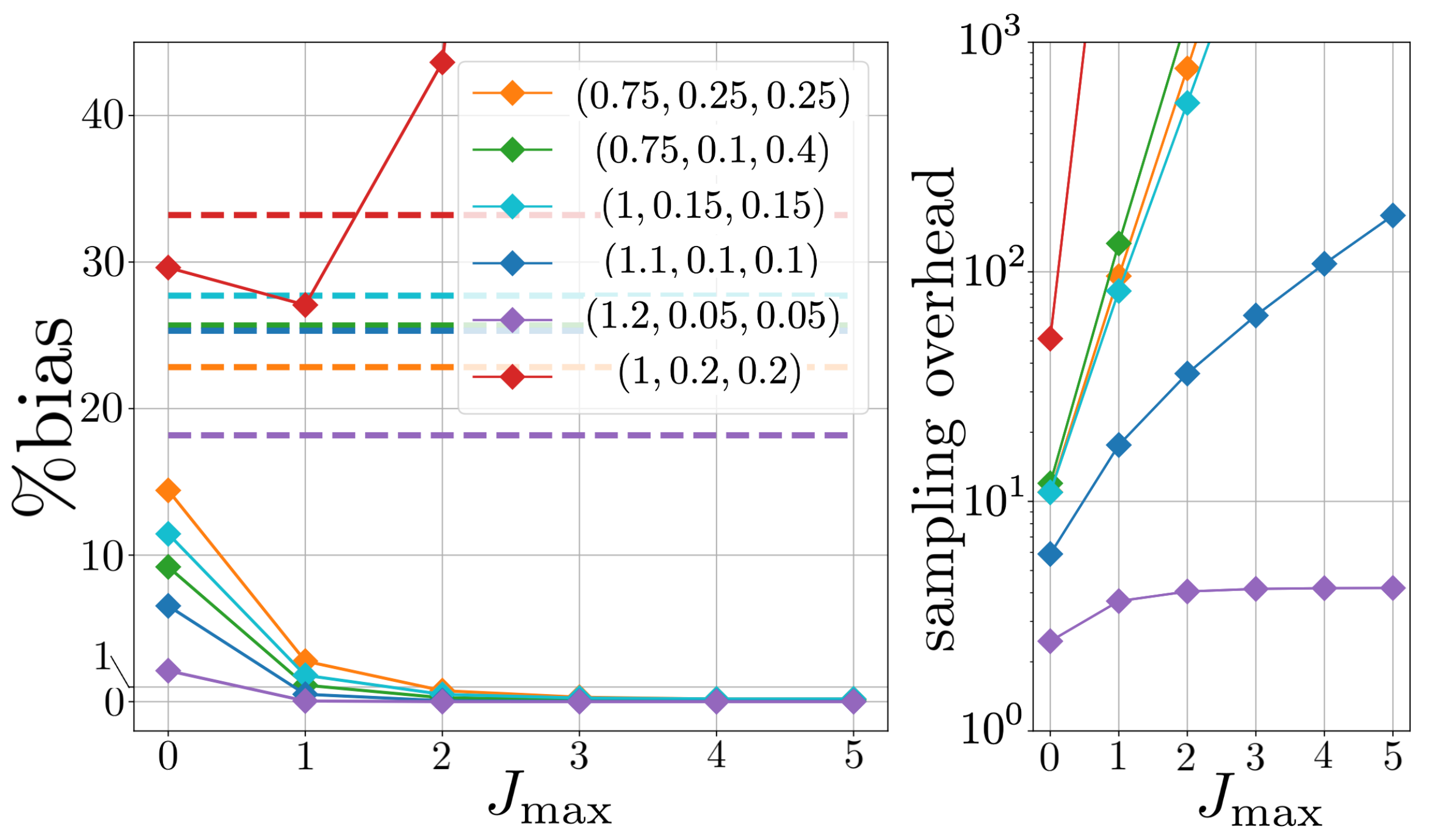}
    \caption{On the left, we plot the bias of $\mathbb{E}[\bar{O}_{\textrm{mit}}] =F(\rho_0, \rho_{\textrm{mit}})$ as a function of $J_{\textrm{max}}$ for the set $\bm{J}^{(\textrm{local})}(J_{\textrm{max}})$. Each curve is labelled by $(r_0, \gamma_1, \gamma_2)$. The dashed lines give the noisy expectation values, $\expval{O}_{\textrm{noisy}} = F(\rho_0, \Lambda_{\bm{\gamma}} [\rho_0])$.
    With the exception of the red curve, we see a rapid convergence to the ideal expectation value as $J_{\textrm{max}}$ grows, and even the zero-photon subtraction channel yields a drastic improvement over the unmitigated results. Subtracting just up to one photon from each mode is sufficient to achieve $\mathcal{O}(1\%)$ bias and in the small loss regime this is improved to $\mathcal{O}(0.1\%)$.
    The red curve diverges because $\Omega_{\bm{j}}$ increases as $j_1$ and $j_2$ increase in the regime where $\bm{j}$ photon losses are non-negligible - this behaviour can be calculated a priori and hence can be avoided by redefining $r_0$.
    On the right, we plot the approximate sampling overhead, $\sum_{\bm{j} \in \bm{J}} |\Omega_{\bm{j}}|^2 / \tilde{p}_{\bm{j}}$ as function of $J_{\textrm{max}}$ under the same choice of parameters. We see a clear exponential growth with respect to $J_{\textrm{max}}$ with a gradient that has a larger dependence on the strength of noise than initial squeezing parameter. For each $(r_0, \bm{\gamma})$, the beam splitter parameters $\bm{\mu} = (\mu, \mu)$ \footnote{Beam splitter parameters are chosen to the nearest $0.005$.} minimise sampling overhead when $J_{\textrm{max}} = 1$. 
    The squeezing parameter associated with $\hat{\mathcal{G}}_{\bm{g_0}} [\rho_0]$ and $\hat{\mathcal{G}}_{\bm{g_\mu}} [\rho_0]$ for each curve, (orange, green, \dots, purple, red), are given by: $(1.245, 1.746)$, $(1.310, 1.734)$, $(1.452, 1.924)$, $(1.420, 1.794)$, $(1.365, 1.552)$ and $(1.856, 2.335)$. These are all accessible with current devices. One can decrease the necessary amplification $\bm{g}_{\bm{\mu}}$ by accepting a larger sampling overhead. 
    }
    \label{fig:TMSV_bias_various_parameters}
\end{figure}

One can show that the quasi-probability distribution $\Omega_{\bm{j}}$ is given by
\begin{align}
\begin{split}
    &\Omega_{\bm{j}} (\ketbra{S_2(r_0)}) = \frac{(-\gamma_1)^{j_1} (-\gamma_2)^{j_2}}{\cosh[2](r_0)} \\ 
    &\times \sum_{n = \textrm{max}[j_1, j_2]}^\infty {n \choose j_1} {n \choose j_2} \left(\frac{\tanh[2](r_0)}{(1 - \gamma_1) (1 - \gamma_2)}\right)^n .
\end{split}
\end{align}
Again, this converges when $\tanh(r_0) / \sqrt{(1 - \gamma_1) (1 - \gamma_2)} < 1$. For fixed choices of $r_0$ and $\bm{\gamma}$, if increasing $j_1$ and/or $j_2$ leads to $\Omega_{\bm{j}}$ increasing, then the quasi-probability weight is pushed towards the channels $\mathcal{E}_{\bm{j}}^{(M)}$ that can not be implemented. This leads to mitigated expectation values that can look divergent in $j_1, j_2$ and places further constraints on what values $r_0$ and $\bm{\gamma}$ can take. An example of this is seen in the red curve of Fig.~\ref{fig:TMSV_bias_various_parameters}. This has parameters $(r_0, \gamma_1, \gamma_2) = (1, 0.2, 0.2)$ and starts rapidly diverging away from the ideal result - in principle, we know that $\expval{O}_{\textrm{ideal}} = \sum_{\bm{j}} \Omega_{\bm{j}} \expval{O}_{\bm{j}}$ but in practise, any numerical cutoff we tested led to divergent behaviour. Numerically, we found that keeping $\tanh(r_0)/\sqrt{(1 - \gamma_1)(1 - \gamma_2)} \lesssim 0.9$ tends to ensure that, for the $\bm{j}$ we are interested in, the expectation value converges to the ideal result. This limit can be increased slightly if $\gamma_1 \ne \gamma_2$ and $\gamma_i \ll 1$. 

Interestingly, $\Omega_{\bm{j}}$ can initially decrease with $\bm{j}$ for ``small'' values of $j_1$ and $j_2$ before beginning to increase with $\bm{j}$ for ``large'' values of $j_1$ and $j_2$ in such a way that these higher photon subtraction terms dominate the quasi-probability. 
Fortunately, this effect is negated by the fact that within the Kraus representation of $\Lambda_{\bm{\gamma}}[\rho_0]$, the terms corresponding to high photon subtraction terms are themselves vanishingly small and therefore the net effect is no appreciable impact on the expectation value.

In Fig.~\ref{fig:TMSV_bias_various_parameters}, we plot the expected percentage bias of the mitigated estimator, $F(\rho_0, \rho_{\textrm{mit}})$, as well as the sampling overhead for various choices of $(r_0, \bm{\gamma}, \bm{\mu})$. We choose the set $\bm{J} = \bm{J}^{(\textrm{local})}$ which is defined in terms of a parameter $J_{\textrm{max}}$ by 
\begin{equation}\label{appendix_def_J_local}
    \bm{J}^{(\textrm{local})} (J_{\textrm{max}}) = \lbrace (j_1, j_2) \,|\, j_1 \le J_{\textrm{max}}, \, j_2 \le J_{\textrm{max}} \rbrace   \, .
\end{equation}
In other words, $J_{\textrm{max}}$ labels the maximum number of photons subtracted \emph{locally} in each mode. We note in passing that the choice of $\bm{J}$ can have a large impact on the bias - for example, the set $\bm{J} = \bm{J}^{(\textrm{global})}(J_{\textrm{max}}) := \{(j_1, j_2)|j_1 + j_2 \le J_{\textrm{max}}\}$ is often outperformed by $\bm{J}^{(\textrm{local})}$, even in cases where $\bm{J}^{(\textrm{local})} (J_{\textrm{max}, 1}) \subset \bm{J}^{(\textrm{global})} (J_{\textrm{max}, 2})$.

\begin{figure}
    \centering
    \includegraphics[width = 0.48\textwidth]{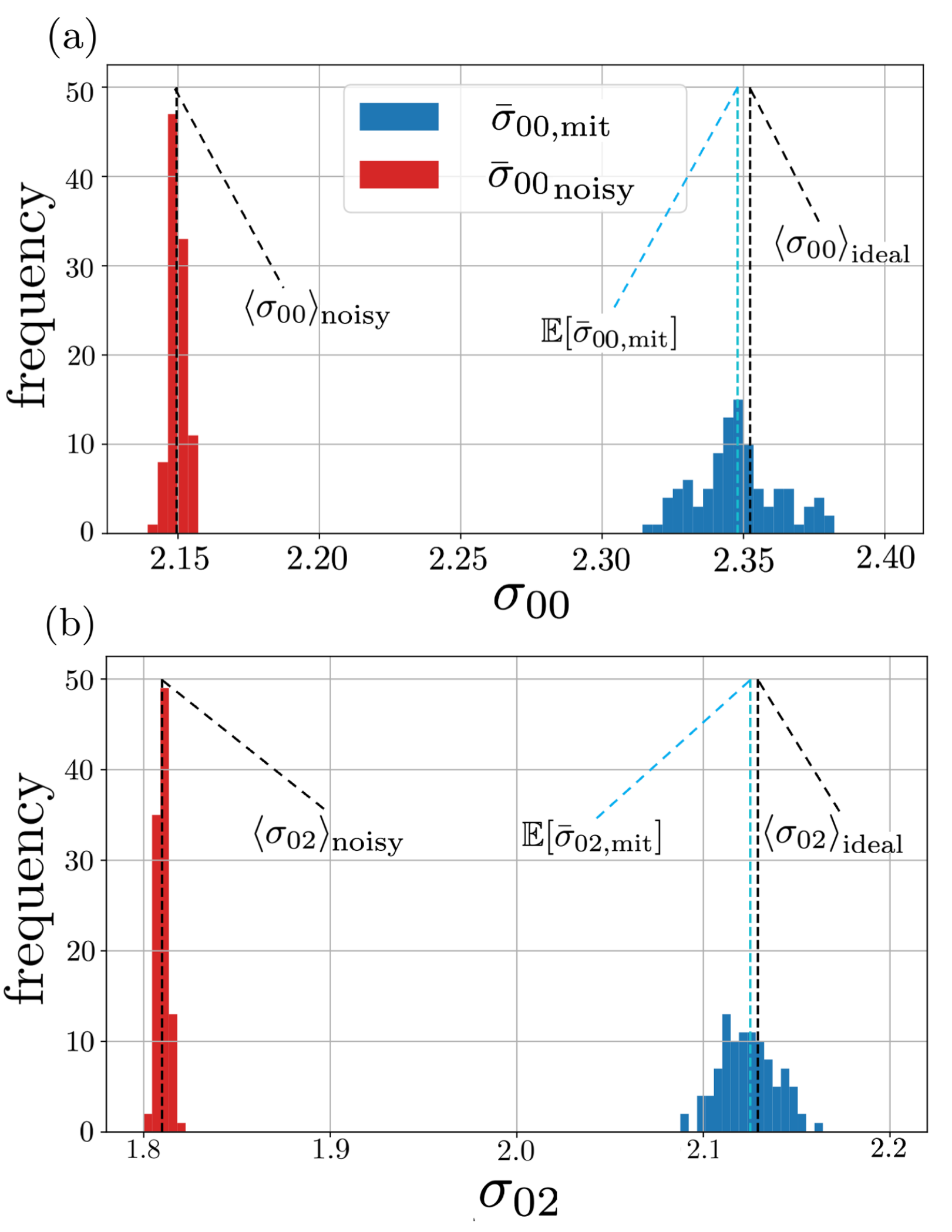}
    \caption{Two histograms illustrating the distribution of results from $100$ simulated experiments in which (a) $\sigma_{00}$ and (b) $\sigma_{02}$ were measured. 
    The initial state, $\ket{S_2(0.75)}$, was subject to losses stochastically chosen from a discretised Gaussian distribution centered on $\bm{\bar{\gamma}} = (0.15, 0.15)$ with standard deviation $\bm{\sigma_{\bm{\gamma}}} = 0.1 \bm{\bar{\gamma}}$ in each shot.
    $1,\!000,\!000$ shots were measured in each experiment to give an estimate of $\sigma_{k \ell}$ and the photon subtraction beam splitters were parametrised by $\bm{\mu} = (0.105, 0.105)$. The red bars show the unmitigated results, the blue bars show the mitigated results and the labelled vertical dashed lines give the ideal, noisy and mitigated expectation values. The percentage bias decreases from $(8.6\%, 15.0\%)$ to $(0.19\%, 0.19\%)$ while the sampling overhead is a modest $\approx 4.4$ and hence our mitigation scheme gives a vast improvement over the unmitigated results, as clearly demonstrated by the histograms.}
    \label{fig:covariance_matrix_histogram}
\end{figure}

The results show a fast convergence to the ideal expectation value for all the results except the red curve, where $(r_0, \gamma_1, \gamma_2) = (1, 0.2, 0.2)$. This red diverging curve is explained by the fact that $\Omega_{\bm{j}}$ increases with $\bm{j}$ - the factor $\tanh(r_0) / \sqrt{(1 - \gamma_1) (1 - \gamma_2)} \approx 0.952 > 0.9$. Decreasing $\gamma_i$ from = $0.2$ to $0.15$ changes the factor to $\tanh(r_0) / \sqrt{(1 - \gamma_1) (1 - \gamma_2)} \approx 0.896 < 0.9$ which does converge correctly as discussed above. The sampling overhead is also plotted. We see that it grows exponentially with $J_{\textrm{max}}$ and the gradient is most significantly impacted by the level of noise $\bm{\gamma}$.
\\
\\
We will now briefly discuss the covariance matrix before using it as the observable in simulations plotted in Fig~\ref{fig:covariance_matrix_histogram}. For further details on the covariance matrix, refer to \cite{serafini2017quantum, RevModPhys.84.621}.

Two-mode continuous variable systems can be described by the quadrature operators $\hat{x}_i=\frac{1}{\sqrt{2}} (\hat{a}_i + \hat{a}_i^\dag)$ and $\hat{p}_i = \frac{1}{i\sqrt{2}} (\hat{a}_i - \hat{a}_i^\dag)$ for $i \in \{1, 2\}$. These can be collected into single $4$-component vector $\bm{\hat{r}} = (\hat{x}_1, \hat{p}_1, \hat{x}_2, \hat{p}_2)^T$. The covariance matrix $\bm{\sigma}$ has matrix elements \cite{serafini2017quantum}
\begin{equation}
    \sigma_{k \ell} = \expval{\{\hat{r}_k, \hat{r}_\ell\}} - 2\expval{\hat{r}_k} \expval{\hat{r}_\ell}
\end{equation}
where $\{A, B\} := AB + BA$ is the anti-commutator. States that admit a Gaussian Wigner function are called Gaussian states and they can be fully characterised by $\expval{\bm{\hat{r}}}$ and $\bm{\sigma}$ \cite{serafini2017quantum}. TMSV states and TMSV states that have been subject to loss are both examples of Gaussian states with $\expval{\bm{\hat{r}}} = 0$ and are hence fully characterised by their respective covariance matrices - by measuring $\bm{\sigma}$, we are essentially carrying out state tomography on $\ketbra{S_2(r)}$ and $\Lambda_{\bm{\gamma}}^{(2)}[\ketbra{S_2(r)}]$.

For both of these states, of the $16$ matrix elements in $\bm{\sigma}$, it can be shown that (up to a sign) there are only $2$ unique non-zero terms:
\begin{equation}
    \bm{\sigma}(\ketbra{S_2(r)}) = 
    \begin{pmatrix}
    \cosh(2r) \bm{I} & \sinh(2r) \bm{Z} \\
    \sinh(2r) \bm{Z} & \cosh(2r) \bm{I}
    \end{pmatrix}
\end{equation}
where $\bm{I}$ is the $2\times 2$ identity matrix and $\bm{Z}$ is the $Z$-Pauli matrix. Hence if one measures $\sigma_{00}$ and $\sigma_{02}$, one has fully characterised the state.

In Fig.~\ref{fig:covariance_matrix_histogram}, we simulate measuring $\sigma_{00}$ and $\sigma_{02}$ from a TMSV state with parameter $r_0 = 0.75$ subject to losses centred on $\bm{\bar{\gamma}} = (0.15, 0.15)$. In each shot, the true photon loss parameter is stochastically taken from a discretized Gaussian distribution of standard deviation $0.1 \bm{\bar{\gamma}}$. To carry out the photon subtractions, two beam splitters of reflectivities $\bm{\mu} = (0.105, 0.105)$ were chosen \footnote{This choice of $\bm{\mu}$ was selected because, to the nearest $0.005$, it optimised the sampling overhead.}. This gave an amplified squeezing parameter of $r_{\textrm{amp by }\bm{g}_{\bm{\mu}}} \approx 1.2041$ (compare this with the minimum necessary squeezing for mitigation, $r_{\textrm{amp by }\bm{g_0}} \approx 0.9666$). We simulated 100 experiments with 1,000,000 shots per experiment and chose $\bm{J} = \bm{J}^{(\textrm{local})}(3)$.

Our error cancellation protocol successfully recovers the noise-free results to within a good approximation, and vastly outperforms the noisy results. The ideal expectation value recovery takes place in spite of a fluctuating $\bm{\gamma}$ which is evidence that our scheme is robust in realistic settings.

Because $\sigma_{00}$ and $\sigma_{02}$ are sufficient to fully characterise $\ket{S_2(r_0)}$, the recovery of the ideal expectation value means we are able to conduct tomography on a Gaussian state that has passed through an unwanted photon loss channel. One can generalise this to tomography of arbitrary state that have undergone unwanted losses by measuring, for example, the Wigner function as an expectation value \cite{PhysRevA.15.449}.

\subsubsection{Entangled Coherent States}
Consider now the two-mode entangled coherent states given by
\begin{equation}
    \ket{\psi_{\pm} (\alpha, \beta)} = \frac{1}{\mathcal{B}_{\pm} (\alpha, \beta)} \left( \ket{\alpha}\ket{\beta} \pm \ket{-\alpha}\ket{-\beta} \right) \,,
\end{equation}
where $\mathcal{B}_\pm (\alpha, \beta) = \sqrt{2 (1 \pm \exp[-2(|\alpha|^2 + |\beta|^2)])}$ is the necessary normalisation. We define $\rho_{\pm}(\alpha, \beta) = \ketbra{\psi_\pm (\alpha, \beta)}$.
If the system is subject to losses $\bm{\gamma} = (\gamma_1, \gamma_2)$, then it can be shown that the operator-sum representation of the inverse map acting on $\rho_-(\alpha, \alpha)$ can be rewritten using only two terms, i.e.,
\begin{align}
\begin{split}
    \Lambda_{\bm{\gamma}}^{-1 (M)} [\rho_- (\alpha, \alpha)] = \tilde{W}_{\textrm{even}} \, \rho_-(\tilde{\alpha}_1, \tilde{\alpha}_2) + \\
    \tilde{W}_{\textrm{odd}} \, \rho_+ (\tilde{\alpha}_1, \tilde{\alpha}_2) \,.
\end{split}
\end{align}
Here, $\tilde{\alpha}_i = \alpha / \sqrt{1 - \gamma_i}$ and the coefficients $\tilde{W}_{\textrm{even/odd}}$ are given by
\begin{align}
\begin{split}
    \tilde{W}_{\textrm{even}} = e^{\gamma_1 |\tilde{\alpha}_1|^2 + \gamma_2 |\tilde{\alpha}_2|^2} \left|\frac{\mathcal{B}_-(\tilde{\alpha}_1, \tilde{\alpha}_2)}{\mathcal{B}_-(\alpha, \alpha)} \right|^2 \\
    \Big( \cosh(\gamma_1 |\tilde{\alpha}_1|^2)\cosh(\gamma_2 |\tilde{\alpha}_2|^2)  \\
    + \sinh(\gamma_1 |\tilde{\alpha}_1|^2) \sinh(\gamma_2 |\tilde{\alpha}_2|^2) \Big) \, ,
\end{split}
\end{align}
\begin{align}
\begin{split}
    \tilde{W}_{\textrm{odd}} = - e^{\gamma_1 |\tilde{\alpha}_1|^2 + \gamma_2 |\tilde{\alpha}_2|^2} \left|\frac{\mathcal{B}_+(\tilde{\alpha}_1, \tilde{\alpha}_2)}{\mathcal{B}_-(\alpha, \alpha)} \right|^2 \\
    \Big( \cosh(\gamma_1 |\tilde{\alpha}_1|^2)\sinh(\gamma_2 |\tilde{\alpha}_2|^2)  \\
    + \sinh(\gamma_1 |\tilde{\alpha}_1|^2) \cosh(\gamma_2 |\tilde{\alpha}_2|^2) \Big) \, .
\end{split}
\end{align}
Notice the key negative sign on $\tilde{W}_{\textrm{odd}}$. If one can produce the entangled coherent states $\ket{\psi_{\pm}(\tilde{\alpha}_1, \tilde{\alpha}_2)}$, then the Monte-Carlo approach from Sec.~\ref{appendix_subsec:monte-carlo_approach_multiple-modes} can be used to give an unbiased estimator of $\expval{O}_{\textrm{ideal}}$. The sampling overhead is approximately given by $S^2$, where $S = |\tilde{W}_{\textrm{even}}| + |\tilde{W}_{\textrm{odd}}|$.

\begin{figure}
    \centering
    \includegraphics[width = 0.49\textwidth]{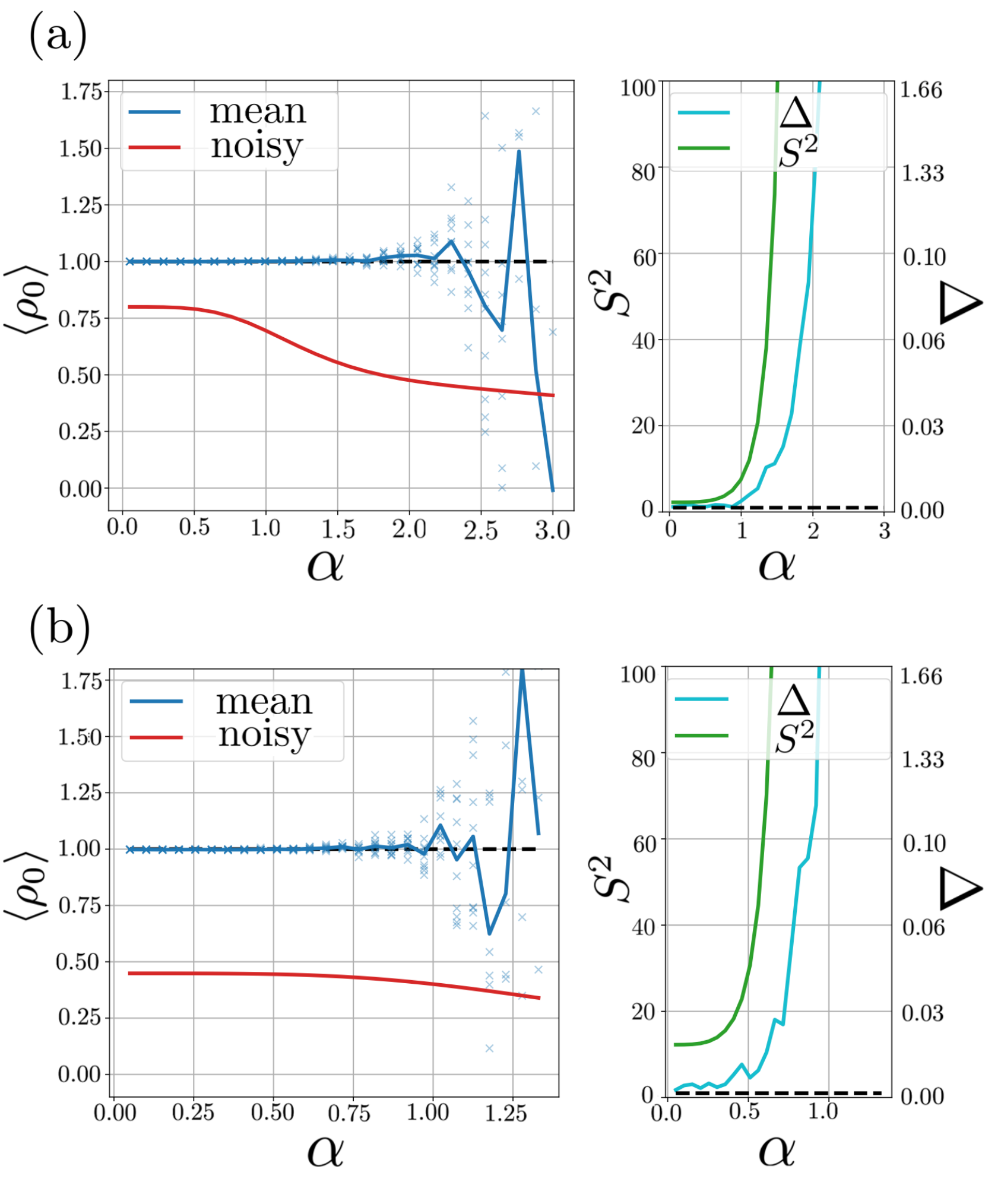}
    \caption{Simulation of $\expval{\rho_0} = F(\rho_0, \rho_{\textrm{mit}})$ against $\alpha$ for the entangled coherent states $\propto \ket{\alpha, \alpha} - \ket{-\alpha, -\alpha}$ subject to losses using our mitigation scheme. 
    In each shot, the loss parameter was stochastically selected from a discretised Gaussian distribution centered on $\bar{\bm{\gamma}}$ with width $\sigma_{\bm{\gamma}} = 0.1 \bar{\bm{\gamma}}$.
    In (a), the average losses are $\bar{\bm{\gamma}} = (0.2, 0.2)$ while in (b) the average losses are $\bar{\bm{\gamma}} = (0.5, 0.6)$.
    For each value of $\alpha$, we conduct $10$ experiments with $1,\!000,\!000$ shots per experiment - each experiment is marked by a tick on the graph and the blue curve is the average over the $10$ experiments. We see a clear recovery of the ideal expectation value for $\alpha \lesssim 2.25$ in (a) and $\alpha \lesssim 1$ in (b). Beyond these points, the large sampling requirements make individual experiments unreliable and can give fidelities that drastically exceed $1$. The graphs on the right plot $S^2 \approx$ sampling overhead and $\Delta$, which is defined as difference between the maximum and minimum values of $\bar{O}_{\textrm{mit}}$ for each $\alpha$. We see a clear relationship between the two which indicates that the sampling overhead is accurately characterising the shot noise.}
    \label{fig:ECS_fidelity_estimate}
\end{figure}

In Fig.~\ref{fig:ECS_fidelity_estimate}, we plot simulated experiments in which the fidelity between $\rho_0$ and $\rho_{\textrm{mit}}$ was measured as an expectation value.
As there is no cutoff in the inverse channel (i.e., Eq.~\eqref{eq:mm_physically_sum_inverse}) with this implementation, this provides an unbiased estimate for all values of $\alpha$, $\beta$ and $\gamma$. On the right, we plot $S^2 \sim \textrm{sampling overhead}$, and the difference between the maximum and minimum values of $\bar{O}_{\textrm{mit}}$ for a given $\alpha$, labelled $\Delta$. In both (a) and (b), the sampling overhead grows exponentially with $\alpha$, which leads to a drastic increase in the shot-noise. Beyond some critical coherent state magnitude, $\alpha_{\textrm{crit}}$, the shot noise dominates and makes individual experiments unreliable. Nonetheless, there is a large and interesting parameter regime in which losses can be mitigated, with just a modest number of shots - for example, the ideal result is consistently recovered for an entangled coherent state of magnitude $\le 1$ even when total losses exceed $50\%$.

\begin{figure}
    \centering
    \includegraphics[width = 0.49\textwidth]{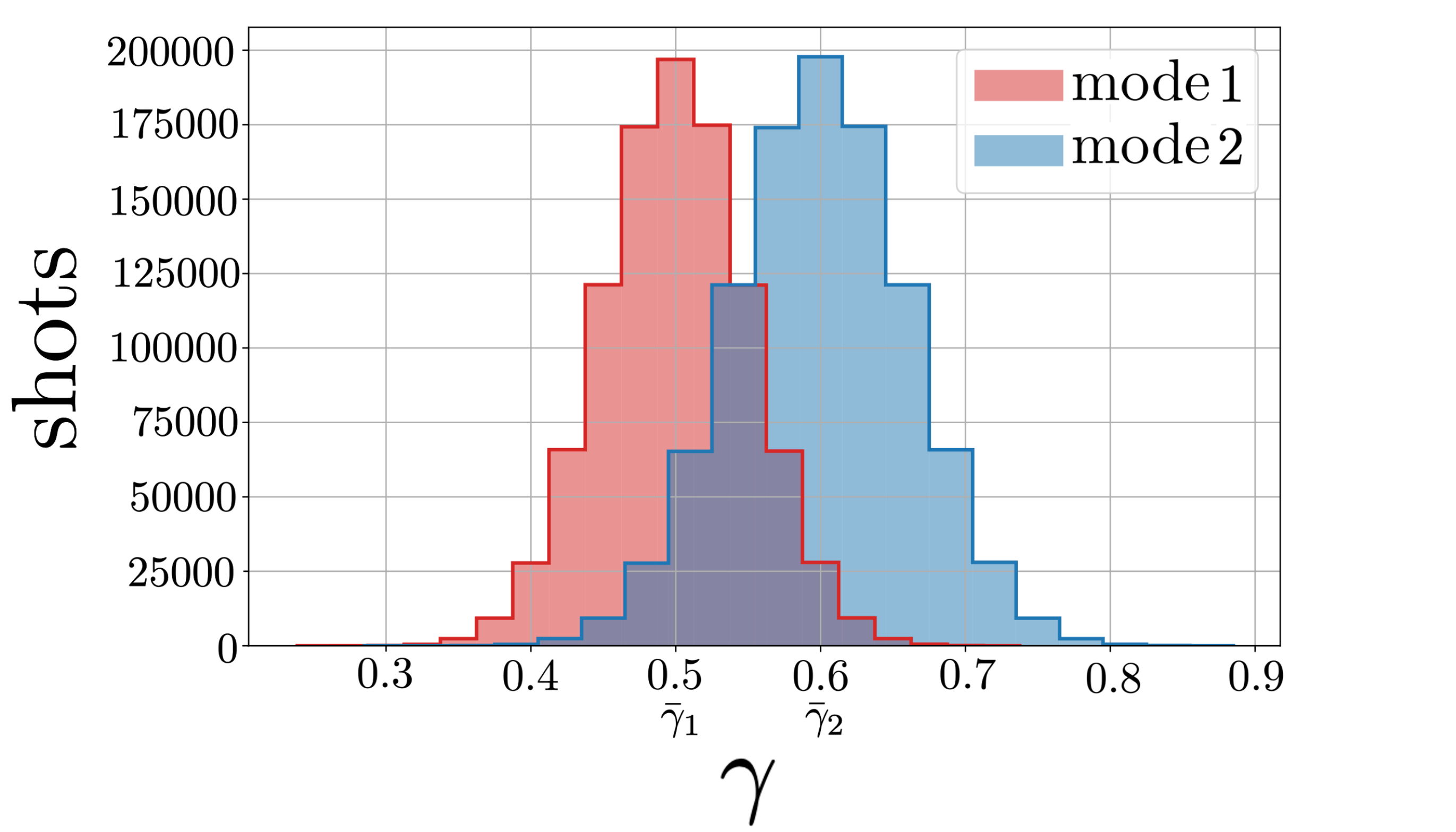}
    \caption{A histogram representative of the $\gamma$ values sampled from in Fig.~\ref{fig:ECS_fidelity_estimate} (b) for a particular experiment with $1000000$ shots. The discretized Gaussian distribution depicted here has mean $\bm{\bar{\gamma}} = (0.5, 0.6)$ and standard deviation $\bm{\sigma_{\bar{\gamma}}} = 0.1 \bm{\bar{\gamma}}$. Despite the relatively large spread in possible loss parameters in a particular shot (ranging all the way from $30\%$ to $90\%$ in mode 2), the fidelity recovery seen in Fig.~\ref{fig:ECS_fidelity_estimate} demonstrates that our mitigation scheme is robust against fluctuating losses and hence can be applied successfully in real experiments.
    In each mode, the Gaussian distribution was linearly discretised into 21 possible values. The range of possible $\gamma$ values was truncated at the tails such that the probability of $\gamma$ lying outside the discretised distribution was $\sim 10^{-7}$.}
    \label{fig:ECS_large_varied_loss_histogram}
\end{figure}

In each shot of the simulation, we stochastically select the specific loss parameter $\gamma$ from a discretized Gaussian distribution centred on $\bm{\bar{\gamma}}$ with standard deviation $\bm{\sigma_{\bar{\gamma}}} = 0.1 \bm{\bar{\gamma}}$.
The recovery of the fidelity then indicates that our scheme is robust against a randomly fluctuating loss parameter. In particular, the results in Fig.~\ref{fig:ECS_fidelity_estimate} (b) and Fig.~\ref{fig:ECS_large_varied_loss_histogram} demonstrates that recovery of robust even when the loss parameters are sampled from surprisingly wide distributions. 
In fact, when comparing these results with results in which the loss parameter was equal to $\bar{\bm{\gamma}}$ for all shots, the authors could not distinguish a noticeable difference.

\section{\label{sec:discussion}Discussion}
In this paper, we have proposed a quantum error mitigation protocol applicable to discrete or continuous variable photonic systems subject to excitation losses, inspired by DV probabilistic error cancellation.
We have analytically inverted the photon loss channel and expressed it using a quasi-probability distribution over physically realisable CP channels. These CP channels involve both a noiseless amplification and multi-photon subtractions and we have proposed viable experimental protocols for implementing both these operations.
Following this, we have then shown how the noisy dynamics can be run and the measurement outcomes combined to construct an estimator of $\expval{O}_{\textrm{ideal}}$ with a bias that can be made arbitrarily small. Furthermore, we have presented a second approach based on Monte-Carlo estimation for certain classes of input states---single photon states, cat states and entangled coherent states---and proven that this produces an unbiased estimator of $\expval{O}_{\textrm{ideal}}$.
In both cases, an unavoidable sampling overhead is introduced. Similar to DV PEC, where the sampling overhead scales exponentially with qubit number and circuit depth, here it scales exponentially with the number of modes and noise strength. A new behaviour observed is the exponential dependence on the \emph{energy} of the system through either the squeezing parameter for squeezed states, or the coherent state magnitude for cat states and entangled coherent states.
Nevertheless, the high shot rate which is characteristic of photonic platforms can ease the impact of this cost.

We have assessed the protocol's performance via several examples.
The squeezed state example displays how the mitigation scheme will typically be employed, and shows that the percentage bias can be made less than $1\%$ within currently experimentally accessible parameter regimes. The protocol was shown to be robust against imperfect characterisation of the loss parameter and a fluctuating loss parameter.
On the other hand, the cat state example illustrates a special case where one can exploit the state's property of being an eigenstate of $\hat{a}^2$ to make the estimate's bias vanish entirely. This Monte-Carlo approach has a better scaling to multiple modes and can be implemented fairly straightforwardly for single photon states, cat states and entangled coherent states.

We hope this paper encourages new work on QEM in CV systems, whether that be adapting existing DV techniques to the CV setting or finding entirely new error mitigation strategies.
A pertinent research direction involves extending this protocol to include dynamic circuits with mid-computation projective measurements - this will be particularly relevant due to the necessity of mid-circuit measurements for universal linear optical quantum computing.
This protocol would also work particularly well with states that are naturally robust against dephasing errors in noise-biased situations. Finally, we hope this work inspires experimental work in this area.
\\


\begin{acknowledgments}
    The authors would like to thank Zhenghao Li, Ewan Mer and Steve Kolthammer for insightful discussions. 
    This project was supported by the UK EPSRC through EP/Y004752/1 and EP/W032643/1.
    G.B. is part of the AppQInfo MSCA ITN which received funding from the European Union’s Horizon 2020 research and innovation programme under the Marie Sklodowska-Curie grant agreement No 956071.
    H.K. is supported by the KIAS Individual Grant No. CG085301 at Korea Institute for Advanced Study and National Research Foundation of Korea (2023M3K5A109480511, 2023M3K5A1094813).
\end{acknowledgments}

\bibliography{biblio}

\appendix


\section{\label{appendix_sec:useful_commutation_relation}Useful Commutation Relation}
In this section, we prove the following identity:
\begin{equation}
\label{appendix_eq:useful_commutation_relation}
    g^{\hat{n}} \hat{a}^j = g^{-j} \hat{a}^j g^{\hat{n}}. 
\end{equation}
We start by showing via induction that the commutation relation
\begin{equation}
\label{appendix_eq:first_commutation_relation}
    [\hat{a}^j, \hat{n}] = j \hat{a}^j
\end{equation}
holds for all $j$. 
The $j = 0$ case is trivial and the $j = 1$ case can be seen by using the canonical commutation relation $[\hat{a}, \hat{a}^\dag] = \mathbb{I}$. 
The induction step reads
\begin{align}
\begin{split}
    [\hat{a}^{j + 1}, \hat{n}] &= \hat{a} \hat{a}^j \hat{n} - \hat{n} \hat{a}^j \hat{a} \\
    &= \hat{a}^{j + 1} \hat{n} -\hat{a}^j \hat{n} \hat{a} + [\hat{a}^j, \hat{n}] \hat{a} \\
    &= \hat{a}^{j + 1} \hat{n} - \hat{a}^j \left(\hat{a} \hat{n} - \hat{a} \right) + j \hat{a}^{j + 1} \\
    &= (j + 1) \hat{a}^{j + 1},
\end{split}
\end{align}
concluding our proof of Eq.~\eqref{appendix_eq:first_commutation_relation}.
Next, we prove that
\begin{equation}
\label{appendix_eq:second_commutation_relation}
    [\hat{a}^j, \hat{n}^k] = \hat{a}^j \left(\hat{n}^k - (\hat{n} - j \mathbb{I})^k\right)
\end{equation}
for all $j$ and $k$ by induction. For $k = 0$ this trivially holds for arbitrary $j$. For $k = 1$, the right hand side reduces to $j \hat{a}^j$ - we have just proven that $[\hat{a}^j, \hat{n}] = j\hat{a}^j$ for arbitrary $j$. Now, we show that if Eq.~\eqref{appendix_eq:second_commutation_relation} holds for arbitrary $j$ and a particular $k$, then it must also hold for arbitrary $j$ and $k + 1$:
\begin{align}
\begin{split}
    [\hat{a}^j, \hat{n}^{k + 1}] &= \hat{a}^j \hat{n}^{k + 1} - \hat{n} \hat{n}^k \hat{a}^j \\
    &= \hat{a}^j \hat{n}^{k + 1} - \hat{n} \left(\hat{a}^j \hat{n}^k - [\hat{a}^j, \hat{n}^k]\right) \\
    &= \hat{a}^j \hat{n}^{k + 1} - \hat{n}\hat{a}^j (\hat{n} - j\mathbb{I})^k  \\
    &= \hat{a}^j \hat{n}^{k + 1} - \hat{a}^j (\hat{n} - j\mathbb{I})^{k + 1} \\
    &= \hat{a}^j \left(\hat{n}^{k + 1} - (\hat{n} - j \mathbb{I})^{k + 1}\right)
\end{split}
\end{align}
By induction, we have therefore shown that Eq.~\eqref{appendix_eq:second_commutation_relation} holds for arbitrary non-negative integers $j$ and $k$.

We now Taylor expand $g^{\hat{n}} = \sum_{k = 0}^\infty \frac{\ln(g)^k}{k!} \hat{n}^k$ and look at $g^{\hat{n}} \hat{a}^j$:
\begin{align}
\begin{split}
    g^{\hat{n}} \hat{a}^j &= \sum_{k = 0}^\infty \frac{\ln(g)^k}{k!} \hat{n}^k \hat{a}^j \\
    &= \hat{a}^j \sum_{k = 0}^\infty \frac{\ln(g)^k}{k!} (\hat{n} - j \mathbb{I})^k \\
    &= \hat{a}^j g^{\hat{n} - j}
\end{split}
\end{align}
Here, we have used the commutation relation in Eq.~\eqref{appendix_eq:second_commutation_relation} in the second line.
This proves the useful relation between $g^{\hat{n}}$ and $\hat{a}^j$ in Eq.~\eqref{appendix_eq:useful_commutation_relation}. This allows us to exchange the order in which the attenuation/amplification and the photon subtractions occur in the loss map/inverse loss map by simply introducing a c-number.


\section{\label{appendix_sec:inverse_excitation_loss_proof}Inverse Photon Loss Map Proof}

In this section, we prove that the $P$-representation and operator-sum representation of the inverse loss map are equivalent. Let $\Lambda_{\gamma}^{-1, (\textrm{op.sum})} [\rho]$ be the operator-sum representation in Eq.~\eqref{eq:inverse_photon_loss_operator_sum_representation}. Via linearity, it is sufficient to show that $\Lambda_{\gamma}^{-1, \textrm{op.sum}} [\ketbra{\alpha}] = \ketbra{g_0 \alpha}$ because then every coherent state in the integrand of Eq.~\eqref{eq:inverse_photon_loss_P_representation} will be correctly amplified.
\begin{align}
\begin{split}
    \Lambda_{\gamma}^{-1, \textrm{op.sum}} [\ketbra{\alpha}] &= \sum_{j = 0}^\infty \frac{(-\gamma)^j}{j!} \hat{a}^j g_0^{\hat{n}} \ketbra{\alpha} g_0^{\hat{n}} (\hat{a}^\dag)^j \\
    &= e^{(g_0^2 - 1) |\alpha|^2} \sum_{j = 0}^\infty \hat{a}^j \ketbra{g_0 \alpha} (\hat{a}^\dag)^j \\
    &= e^{\frac{\gamma}{1 - \gamma} |\alpha|^2} \sum_{j = 0}^\infty \frac{1}{j!} \left(\frac{-\gamma |\alpha|^2}{1 - \gamma}\right)^j \ketbra{g_0 \alpha} \\
    &= \ketbra{g_0 \alpha}.
\end{split}
\end{align}
In the third line we used the fact that $g_0^2 - 1 = \gamma / (1 - \gamma)$. Therefore, applying $\Lambda_{\gamma}^{-1}$ in the operator-sum representation to $\rho$ in the $P$-representation will return $\Lambda_{\gamma}^{-1}$ in the $P$-representation. This proves that these two representations are equivalent.


\section{\label{appendix_sec:photon_subtraction}Photon Subtraction}
In this section, we prove that sending an arbitrary state $\rho$ through a beam splitter with a vacuum in the second entry port and measuring $j$ photons heralds a state conjugated by the $j^{\textrm{th}}$ Kraus operators defined in Eq.~\eqref{eq:photon_loss_kraus_representation}. 

Let the input and output modes be described by the annihilation operators $\hat{a}_i$ and $\hat{b}_i$ respectively, where $i \in \{\textrm{system}, \textrm{ ancilla}\}$. We let the input system state be pure and given by $\ket{\psi} = \sum_{n = 0}^\infty c_n \ket{n} = \sum_{n = 0}^\infty c_n \frac{\hat{a}_S^n}{\sqrt{n!}} \ket{0}$ while the ancilla input state is a vacuum, $\ket{0}$. The order of tensor product is system-ancilla (e.g.; $\ket{0}\ket{0} = \ket{0}_{S} \ket{0}_A$). 

In the Heisenberg picture, a beam splitter of transmissivity $1 - \mu$, described by the unitary $U_{\textrm{BS},\mu}$ defined in Sec.~\ref{subsubsec:photon_subtraction}, will transform $\hat{a}_S^\dag \rightarrow \sqrt{1 - \mu} \, \hat{b}_S^{\dag} - \sqrt{\mu} \, \hat{b}_A^{\dag}$. This acts on $\ket{\psi}\ket{0}$ to produce the state
\begin{align}
\begin{split}
    U_{\textrm{BS}, \mu} \ket{\psi} \ket{0} &= \sum_{n = 0}^\infty c_n  \frac{(\sqrt{1 - \mu} \, \hat{b}_S^{\dag} - \sqrt{\mu} \, \hat{b}_A^{\dag})^n}{n!} \ket{0} \ket{0}, \\
    &= \sum_{n = 0}^\infty \sum_{m = 0}^{n} c_n (\sqrt{1 - \mu})^{n - m} (-\sqrt{\mu})^m \\
    &\quad\quad\quad\quad\quad \times {n \choose m}^{\frac{1}{2}} \ket{n - m} \ket{m},
\end{split}
\end{align}
where ${n \choose m} = n! / (m! (n - m)!)$ is the binomial coefficient.

To calculate the state that is heralded by a measurement of $j$ photons in the ancilla output mode, we first apply $\bra{j}_A$ to the above state:
\begin{align}
\begin{split}
    \bra{j}_A U_{\textrm{BS}, \mu} \ket{\psi}_S \ket{0}_A \\
    = (-1)^j \sqrt{\left(\frac{\mu}{1 - \mu}\right)^j}
    \sum_{n = j}^\infty c_n (1 - \mu)^{\frac{n}{2}} {n \choose j}^{\frac{1}{2}} \ket{n - j} \\
    = (-1)^j \sqrt{\left(\frac{\mu}{1 - \mu}\right)^j} \sum_{n = 0}^\infty c_n (1 - \mu)^{\frac{n}{2}} {n \choose j}^{\frac{1}{2}} \sqrt{\frac{(n - j)!}{n!}} \hat{a}^j \ket{n} \\
    =  (-1)^j \sqrt{\frac{1}{j!} \left(\frac{\mu}{1 - \mu}\right)^j} \hat{a}^j \sum_{n = 0}^\infty c_n (1 - \mu)^{\frac{n}{2}} \ket{n} \\
    = (-1)^j \sqrt{\frac{1}{j!} \left(\frac{\mu}{1 - \mu}\right)^j} \hat{a}^j (1 - \mu)^{\frac{\hat{n}}{2}} \ket{\psi} \\
    = (-1)^j K_j(\mu) \ket{\psi}.
\end{split}
\end{align}
where in the third line we used $\hat{a}^j \ket{n<j} = 0$, in the fourth line ${n \choose j}^{\frac{1}{2}} \sqrt{\frac{(n - j)!}{n!}} = \sqrt{\frac{1}{j!}}$ and in the fifth line $g^n \ket{n} = g^{\hat{n}} \ket{n}$. Using this, we can see that a projective measurement of $j$ photons from the ancilla output port will (up to normalisation) herald the state
\begin{align}
\begin{split}
    \ketbra{\psi} \rightarrow &\Tr[\mathbb{I} \otimes \ketbra{j}) U_{\textrm{BS}, \mu} (\ketbra{\psi}_S \otimes \ketbra{0}_A) U_{\textrm{BS}, \mu}^\dag ] \\
    &= \bra{j}_A U_{\textrm{BS}, \mu} \ket{\psi}_S \ket{0}_A \,\,\, \bra{\psi}_S \bra{0}_A U_{\textrm{BS}, \mu}^\dag \ket{j}_A \\
    &= K_j (\mu) \ketbra{\psi} K_j^\dag (\mu).
\end{split}
\end{align}
This proves the statement for pure states. Via linearity, it can easily be seen that for an arbitrary density operator $\rho = \sum_i \lambda_i \ketbra{\psi_i}$, a measurement of $j$ photons from the ancillary output port will (up to normalisation) herald the state
\begin{equation}
    \rho \rightarrow K_j (\mu) \rho K_j^\dag (\mu).
\end{equation}


\section{Dephasing Errors}\label{dephasing_error_section}
In this section, we briefly discuss bosonic dephasing errors, the impact they have on our loss mitigation scheme and give a new dephasing error mitigation scheme based around the inverse dephasing error map. The single mode pure bosonic dephasing channel, $\Lambda_{D, \gamma_D}$, is defined by
\begin{equation}
    \Lambda_{D, \gamma_D} [\rho] = \sum_{n = 0}^\infty \sum_{m = 0}^\infty e^{-\frac{\gamma_D}{2} (m - n)^2} \matrixel{m}{\rho}{n} \ketbra{m}{n} \,,
\end{equation}
where $\gamma_D \in [0, \infty)$ is a parameter that determines the noise strength. 
There is an exponential damping on the off-diagonal terms that decoheres the state.

This can equivalently be seen to solve the Lindblad master equation
\begin{equation}
\label{ap_eq:dephasing_lindblad_master_equation_representation}
    \frac{d}{dt} \rho (t) = \kappa_D \mathcal{D}[\hat{n}](\rho(t)) \, , 
\end{equation}
where $\mathcal{D}[\hat{n}](\rho) = \hat{n} \rho \hat{n} - \frac{1}{2} \{\hat{n}^2,\, \rho\}$ is the dephasing Lindbladian and $\kappa_D > 0$ is the dephasing rate, if we make the identification $\gamma_D = \kappa_D t$.

The joint loss-dephasing channel, denoted $\Lambda_{LD, \gamma, \gamma_D}$, gives the solution to the Lindblad master equation
\begin{equation}
\label{ap_eq:loss_dephasing_lindblad_master_equation_rep}
    \frac{d}{dt} \rho (t) = \left[\kappa \mathcal{D}[\hat{a}] + \kappa_D \mathcal{D}[\hat{n}]\right] (\rho (t))
\end{equation}
The fact that the two Lindbladians commute means that the joint loss-dephasing channel can be decomposed into a pure loss channel of strength $\gamma$ taking place before or after a pure dephasing channel of strength $\gamma_D$; i.e., $\Lambda_{LD, \gamma, \gamma_D} = \Lambda_\gamma \circ \Lambda_{D, \gamma_D} = \Lambda_{D, \gamma_D} \circ \Lambda_\gamma$. This in turn means that if our system is subject to a joint loss-dephasing channel of strength $(\gamma, \gamma_D)$, then we can use our quantum error cancellation protocol to mitigate against the losses, leaving behind an expectation value that is equivalent to if the system was just subject to the pure dephasing channel of strength $\gamma_D$. In other words, our error cancellation scheme for photon losses is robust against the introduction of a dephasing error channel.

We now briefly introduce a scheme for mitigating against the pure dephasing channel.
A discrete Kraus representation of the dephasing channel is given by
\begin{equation}
    \label{appen_eq:dephasing_channel_kraus_representation}
    \Lambda_{D, \gamma_D} [\rho] = \sum_{j = 0}^\infty \frac{\gamma_D^j}{j!} \hat{n}^j e^{-\frac{\gamma_D}{2} \hat{n}^2}  \rho \, e^{-\frac{\gamma_D}{2} \hat{n}^2} \hat{n}^j \, ,
\end{equation}
where we have explicitly expanded out the Kraus operators $D_j (\gamma_D) = \sqrt{\gamma_D^j / j!} \, \hat{n}^j e^{-\frac{\gamma_D}{2} \hat{n}^2}$. Using this, one can then easily show that the inverse dephasing channel has an operator-sum representation of
\begin{equation}
    \label{appen_eq:inverse_dephasing_channel_operator_sum}
    \Lambda_{D, \gamma_D}^{-1} [\rho] = \sum_{j = 0}^\infty \frac{-\gamma_D^j}{j!} \hat{n}^j e^{\frac{\gamma_D}{2} \hat{n}^2}  \rho \, e^{\frac{\gamma_D}{2} \hat{n}^2} \hat{n}^j \, .
\end{equation}
This corresponds to simply changing the the sign of $\gamma_D \rightarrow -\gamma_D$ in Eq.~\eqref{appen_eq:dephasing_channel_kraus_representation}. 
Similarly to the photon loss inverse map, Eq.~\eqref{appen_eq:inverse_dephasing_channel_operator_sum} can be expressed as a weighted sum (with some negative coefficients) of CP maps.
Consider the channels defined by
\begin{equation}
    \mathcal{E}_{D, j} [\rho] = \frac{1}{\mathcal{N}_{D, j} (\rho)} \hat{n}^j e^{\frac{\gamma_D}{2} \hat{n}^2}  \rho \, e^{\frac{\gamma_D}{2} \hat{n}^2} \hat{n}^j \, ,
\end{equation}
where $\mathcal{N}_{D, j} (\rho)$ is a normalisation coefficient.
The inverse can then be written as linear sum over these channels with coefficients $\omega_{D, j} (\rho) = (-\gamma_D)^j \mathcal{N}_{D, j} (\rho) / j!$. Using this, if one can realise the channels $\mathcal{E}_{D,j}$ then one mitigate against the dephasing channel using similar ideas that have been discussed in this paper for mitigating against the loss channel.

Unfortunately, an experimental realization of the channels $\{\mathcal{E}_{D, j} \}$ seems particularly challenging.
In fact, a realization of $\hat{n}^j$ alone requires post selection that makes scaling beyond a couple of modes unfeasible.
Additionally, unlike the noiseless amplification term present in the inverse excitation loss map, the operator $e^{\frac{\gamma_D}{2} \hat{n}^2}$ may not be easily incorporated in the state preparation stage.
Unless a way around these difficulties is found, the error cancellation protocol presented in this paper is unlikely to be experimentally viable to mitigate against dephasing errors.



\end{document}